\documentclass[singlecolumn,superscriptaddress,showpacs,preprintnumbers,
               amsmath,amssymb]{revtex4}

\usepackage{float}
\usepackage{graphicx}   
\usepackage{dcolumn}    
\usepackage{bm}         
\usepackage{color}      
\usepackage{booktabs}
\usepackage[caption=false]{subfig}

\def\mem#1#2#3{  \left\langle #1 \left\vert  #2 \right\vert #3 \right\rangle   }

\begin{document}


\title{Electromagnetic Heavy Lepton Pair Production in Relativistic Heavy-Ion Collisions}
\author{M.~Y.~\c{S}eng\"{u}l}
\email{melekaurora@yahoo.com}
\noaffiliation{}

\author{M.~C.~G\"{u}\c{c}l\"{u}}
\author{\"{O}.~Mercan}
\author{N.~G.~Karaku\c{s}} 
\affiliation{\.{I}stanbul Technical University, Faculty of Science and Letters,
             34469, \.{I}stanbul,Turkey}
\date{\today}

\begin{abstract}
We calculate the cross sections of electromagnetic productions of muon and tauon pair productions from the ultra-relativistic
heavy ion collisions. Since the Compton wavelengths of muon and tauon are comparable to the radius of the colliding ions, nuclear
form factors play important roles for calculating the cross sections. 
Recent measurement \cite{abrahamyan} indicates that the neutrons are differently distributed
from the protons therefore this affects the cross section of the heavy lepton pair production. In order to see the effects
of the neutron distributions in the nucleus, we used analytical expression of the Fourier transforms of the Wood-Saxon distribution.
Cross section calculations show that Wood-Saxon distribution function is more sensitive to the parameter $R$ compare to the parameter $a$.
\end{abstract}

\maketitle

\section{\label{sec:level1}Introduction}\label{s1}
Colliding beams of heavy ions at energies per nucleon in the
range $10^2 - 10^4$ GeV, corresponding to experiments RHIC at Brookhaven, and LHC at CERN, suggested that intense electromagnetic
fields of these ions can produce relatively large fluxes of muons and tauons. Since mass of the electron (0.511 MeV) is much smaller than
muon (105.66 MeV) and tauon (1784 MeV), and the Compton wavelength of the electron (386 fm) is much larger than muon (1.86 fm) and tauon
(0.11 fm), the cross sections of producing electron pairs are much larger than the heavy leptons. In addition
to this, since the Compton wavelengths of muons and tauons are smaller than the radius of the colliding heavy ions (Au, Pb), 
nucleon \cite{belk1} and nucleus form factors are not negligible. Therefore realistic charge form factors play important roles 
for calculating the cross sections of the heavy lepton pair productions.

In this work, we have used Wood-Saxon nucleus form factor which is also widely used in literature \cite{jent,klusek,baltz1,baltz2}. 
Instead of the monopole approximation, we have an analytical expression of the Wood-Saxon form factor in our equations. 
With the help of this analytical equation, we are able to account for the neutron skin and its effects to our calculations.
The determination of the sizes and shapes of atomic nuclei is one of the important problems in nuclear physics. The rms radius
of the charge distribution in nuclei is well known with the high degree of accuracy achieved by the elastic electron-nucleus
and muon-nucleus scattering experiments. The experimental uncertainties for the nuclear charge radius are about $1\%$ for several
nuclei \cite{angeli}. However, the neutron distribution and its rms radius in nuclei 
is not precisely determined theoretically and experimentally.
Generally, the neutron skin thickness, defined as the neutron-proton rms radius difference in the atomic nucleus,
\begin{eqnarray}
\Delta R_{np} = R_{n} -R_{p} = <r^2>_{n}^{1/2} - <r^2>_{p}^{1/2}
\end{eqnarray}
where the measured charge radius are $<r^2>_{p}^{1/2}=5.43 $ fm for the gold nucleus and $<r^2>_{p}^{1/2}=5.50 $ fm for the lead nucleus
 \cite{jent}.
By using the parity-violating electron scattering (PVES), the recent experiment \cite{abrahamyan} at JLab gives us the value $R_{n}$ as $ 5.78^{+0.16}_{-0.18}$. The difference between the radii of the
neutron and proton distributions is the neutron skin and it is equal to $ R_{skin} = R_{n} - R_{p} = 0.33^{+0.16}_{-0.18}$ fm for 
the lead nucleus with large total error.
Therefore the difference between the distributions of protons and neutrons in the nucleus affects the cross section of the heavy 
lepton pair production.

To explore a surface thickness of the neutron density in a heavy nucleus, we model the neutron density with a Wood Saxon form:
\begin{eqnarray}
\rho_{n}(r) = \rho_{0}/[1 + exp(r-R_{0})/a_{n}]
\end{eqnarray}
where $R_{0}$ and $a_{n}$ are the parameters that determine the shape of the neutron distribution. Although the same parameters are 
precisely determined for the proton distribution in the heavy nucleus, there are large uncertainties for the neutron distribution.
In this work, we study the effect of these parameters to the heavy lepton pair production cross sections.

Although the electron-positron pair production from electromagnetic fields goes back to 1930s \cite{racah,weiz},there are numbers 
of published literature in the 1990s and 2000s \cite{bottch2,guclu1,henc11,best,eichler,baltz3,baltz4}. The main motivation at these period was the RHIC. 
Dilepton production from the central collisions of heavy ions is very important to understand the hadronic interactions. 
However electromagnetic production of dileptons from the peripheral collisions of heavy ions can shield those dileptons produced 
from the hadronic interactions. Therefore it is highly important to understand various properties of the electromagnetically 
produced dileptons \cite{baur2}.

\begin{figure}
\includegraphics[width=8.6cm,height=6.6cm]{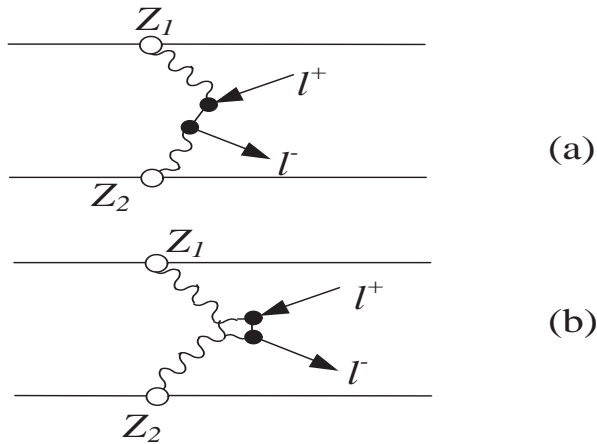} 
\caption{(a) Direct and (b) crossed Feynman diagrams for the heavy lepton pair production in a relativistic heavy ion collision.}
\label{f20} 
\end{figure}

In our previous works \cite{guclu1,mel1,mel2}, we have calculated electron-positron pair production cross sections by 
using the second order Feynman diagrams. In this work, we have modified the equations by including the electromagnetic 
form factors of nuclei in the momentum space. We have used Fermi type charge form factor. We also consider the coherent 
production of heavy lepton pairs in two photon limit for the symmetric collision of two gold nuclei 
at RHIC and two lead nuclei at LHC energies. In Fig.~\ref{f20}, the wiggly lines are the virtual photons that make the 
electromagnetic field of the colliding ions. Here we start with the lowest non-vanishing Feynman diagram which is called 
two-photon diagram. We take the classical limit of the motion of the ions which yields the external field model. Since 
the parameter $Z\alpha \approx 0.6$ is not small, Coulomb and unitarity corrections should be included to obtain 
exact cross sections. This calculations has been done in \cite{henc1,ivanov,lee}, and we will also mention briefly in the 
results section.

In the Formalism section, we briefly explain two photon pair production mechanism. In Calculations and Tables section, 
we study the effects of the form factor that we use in our calculations and we also calculate a variety of cross sections for the 
single-pair process. Finally on the last section, we summarize all the work that we have done in this manuscript.

\section{Formalism}
\begin{figure}
\includegraphics[width=8.6cm,height=6.6cm]{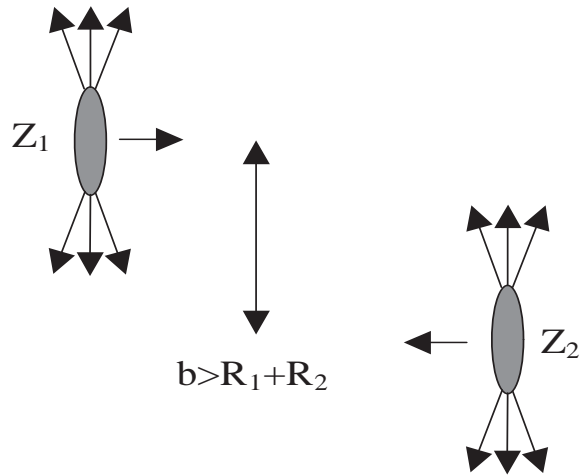} 
\caption{Collision of two heavy ions at ultra-relativistic velocities. Lorentz contracted strong EM fields are produced around the ions.}
\label{f21}
\end{figure}

In this work we will investigate the two-photon mechanism for heavy lepton pair production. In Fig.~\ref{f21}, the electromagnetic
interaction of the colliding ions is shown. Since all the protons in the nucleus act coherently, there is a very strong
electromagnetic field appears around the heavy ions \cite{bottch2}. Lepton pairs are produced from this field which is a cloud of virtual
photons.  Although we consider peripheral collisions of the heavy-ions where the impact parameter $b > R_{1} + R_{2}$ is greater
than the sum of the radii of the colliding nuclei, our cross section expressions include all impact parameter region from zero to infinity.
We consider the electromagnetic production of lepton pairs from the collisions of heavy ions with charge $Z$
\begin{eqnarray}
Z_{1} + Z_{2} \rightarrow Z_{1} + Z_{2} + l^{+}l^{-} 
\end{eqnarray}
where $l^{+}l^{-}$ can be electrons, muons or tauons. The maximum magnitude of the electric field of the ions are
\begin{eqnarray}
E_{max} \simeq \frac{Ze\gamma}{b^{2}}
\end{eqnarray}
where $\gamma$ is the Lorentz factor and $b$ is the impact parameter and $Ze$ is the charge of the fully stripped
heavy ion. For the impact parameters around $b \sim 7.5 fm$ which is very closed to nucleus but still peripheral collisions,
maximum strength of the electric field at the surface $E_{max} \sim 2\times10^{23} V/m$ for the RHIC and $E_{max} \sim 7.14 \times 10^{24} V/m$
for LHC collisions. We can estimate the minimum electric field that can create a lepton pair of mass $2mc^{2}$ as 
\begin{eqnarray}
E_{min} \simeq \frac{m^{2}c^{4}}{e\hbar c}
\end{eqnarray}
where the distance that we used in the above equation is twice the Compton wavelength of the particular lepton\cite{guclu1}. To create $e^{+}e^{-}$,
$\mu^{+}\mu^{-}$ and $\tau^{+}\tau^{-}$ pairs, the minimum electric field needed are $1.32 \times 10^{18} V/m$, 
$5.66 \times 10^{22} V/m$ and $1.62 \times 10^{25} V/m$, respectively. As we see that even at LHC energies, maximum electric field 
produced from the lead ions is below the critical electric field to create tauon pairs. Even 
for the impact parameters around $b \sim 7.5 fm - 30 fm$ regions, maximum electric field is smaller than the electric field that produce
tauon pairs at RHIC and LHC energies. However, in QED there is still a probability to produce heavy leptons even at these energies.

The source currents for producing lepton pairs are the Lorentz-boosted charges of the colliding heavy ions. Electromagnetic
fields in this collisions are very strong that they can pull many lepton pairs from the vacuum. Relativistic heavy-ion collider (RHIC)
at Brookhaven can accelerate the fully stripped gold nuclei at 100 GeV/nucleon energies and large hadron collider (LHC) at CERN can accelerate
the fully stripped lead nuclei up to 3400 GeV/nucleon energies. Therefore electromagnetic fields at this collisions last for a very short time
$\Delta t = b/(\gamma c \beta)$ and the Fourier frequencies could be up to $\omega_{max}\sim \gamma c \beta/b $ \cite{rep2,guclu1}.
We can write the Lagrangian density as sum of the non-interacting fermion Lagrangian $\mathcal{L}_{0}$ and the Lagrangian density
for the coupling of the classical electromagnetic potential to the lepton fields $\mathcal{L}_{int}$
\cite{book1,book2}

\begin{eqnarray}
\mathcal{L}(x) & = & \mathcal{L}_{0}(x) + \mathcal{L}_{int}(x) \nonumber \\
               & = & \overline{\Psi}(x)[\gamma_{\mu}i\partial^{\mu} - m ] \Psi (x) - \overline{\Psi}(x)\gamma_{\mu} \Psi (x) A^{\mu}(x)
\end{eqnarray}
where $ A_{1}^{\mu} \equiv (A_{1}^{0}, \vec{A}_{1})$ is the four-vector potential of the nucleus 1 and the scalar component of the potential
can be written as
\begin{eqnarray}
A_{1}^{0} = -8\pi^{2} Z \gamma^{2}\delta(q_{0}+\beta q_{z})\frac{e^{-i\vec{q}_{\perp}\cdot\vec{b}/2}}{(q_{z}^{2} +\gamma^{2}q_{\perp}^{2})}G_{E}(q^{2})f_{Z}(q^{2})
\end{eqnarray}
and non-zero component of the vector potential $A_{1}^{3}$ is in the longitudinal direction and it can be obtained as
\begin{eqnarray}
A_{1}^{3} = -\beta A_{1}^{0}
\end{eqnarray}
where the transverse components of the vector potentials are $ A_{1}^{1} = A_{1}^{2} = 0 $. The potentials from nucleus 2 can be written 
by the substitutions, $ \vec{b} \rightarrow -\vec{b}$ and $\beta \rightarrow -\beta $. In the above equation $ f_{Z}(q^{2})$ is 
the form factor of a nucleus which gives the momentum distribution of a proton in the nucleus, and $ G_{E}(q^{2}) $ 
is the form factor of a proton which represents electric distribution of the proton\cite{guclu1,bottch1}. As we mention above, we have used 
two parameter Fermi (2pF) function (or Wood-Saxon function) for the charge distribution of the protons:
\begin{eqnarray} \label{ws}
\rho(r) = \dfrac{\rho_{0}}{1 + exp((r-R)/a)}
\end{eqnarray}
where $ R $ is the radius of the nucleus and $a$ is the skin depth or diffuseness parameter. These parameters are obtained by fits to
electron scattering data \cite{barrett} and $\rho_{0}$ is written by the normalization condition.
For symmetric nuclei, the nuclear density for a nucleus that has mass number $A$, a distance $r$
from its center is modeled in literature with a Woods-Saxon distribution as in Eq.~(\ref{ws})
where $\rho_{0}=\frac{0.1694}{A}fm^{-3}$ for Au nucleus and $\rho_{0}=\frac{0.1604}{A}fm^{-3}$ for Pb nucleus. 
The radii of the gold and lead nucleus are equal to $R_{Au} = 6.38$ fm and  $R_{Pb} = 6.62$ fm respectively.
However, recent measurement \cite{abrahamyan} indicates that the neutrons 
are differently distributed from the protons so that the parameters $R$ and $a$ must be modified. 
Therefore this effects the cross section of the heavy lepton pair production \cite{klein}. In order to see 
the effects of the neutron distributions in the nucleus, we need to have analytical expression of the Fourier
transforms of the Wood-Saxon distribution \cite{magnus}:

\begin{eqnarray}
 f_{Z}(q^{2}) & = & \int^{\infty}_{0}\frac{4\:\pi}{q}\rho(r)Sin(qr)dr  \nonumber \\
 & = & \int^{\infty}_{0}\frac{4\:\pi}{qr}\rho(r)Sin(qr)r^{2}dr
\nonumber \\
 & = & \frac{4\pi^{2}\rho_{0}a^{3}}{(qa)^{2}Sinh^{2}(\pi q a)}[\pi q a\:Cosh(\pi q a)Sin(qR)-qRCos(qR)Sinh(\pi q a)]+8\pi \rho_{0}a^{3} 
    \sum^{\infty}_{n=1}(-1)^{n-1}\frac{n e^{-n R/a}}{[n^{2}+(q a)^{2}]^{2}}.
\end{eqnarray}
\begin{figure}
\includegraphics[width=8.6cm,height=6.6cm]{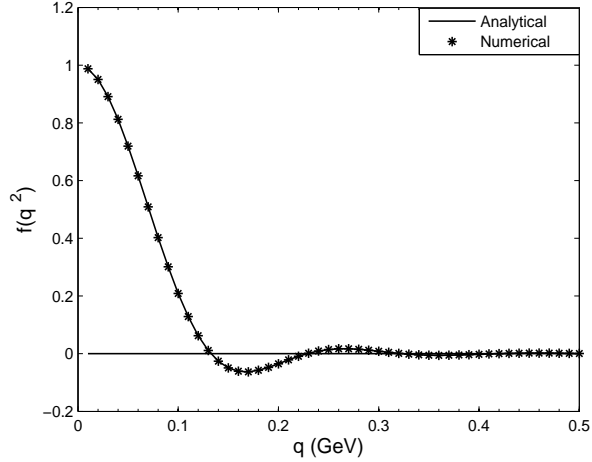} 
\caption{The electromagnetic form factor for gold. The solid line is the analytic form of Woods-Saxon distribution and the star line is the numerical calculation of the same equation.}
\label{f31}  
\end{figure}
The parameters $R$ and $a$ are explicitly shown in this analytical equation and it is plotted in Fig.~\ref{f31} together with the numerical result 
from Fourier transformation of a Woods-Saxon distribution. The agreement is excellent. Since the contribution of the last term 
in above equation is very small compare to the first term, it is safely neglected.  By changing the parameters $R$ and $a$ 
we can estimate the sensitivity of the nucleon distributions in the nucleus. 

On the other hand, for the electric distribution function of the proton, we can write the dipole
form factor for the proton as
\begin{eqnarray}
G_{E}(q^{2}) = (\dfrac{\Lambda^{2}}{\Lambda^{2}+ q^{2}})^{2}
\end{eqnarray}
where the value of $\Lambda^{2}$ is 0.71 $GeV^{2}$\cite{bottch1}. We have observed that the effects of this
distribution is negligible.
Finally, the total electromagnetic potential
is sum of the nucleus 1 and 2
\begin{eqnarray}
A^{\mu} = A_{1}^{\mu} + A_{2}^{\mu}.
\end{eqnarray}

To calculate the total cross section for the heavy lepton pair production, we first represent the direct term,

\begin{widetext}
\begin{eqnarray} \label{d11}
   \mem{\Psi^{(+)}_k}{\,S_{ab}\,}{\Psi^{(-)}_q} & = &
   i\,\sum_{p}\sum_{s} \int^{\infty}_{-\infty} \frac{d\omega}{2\pi} \:
   \frac{\mem{\Psi^{(+)}_k}{V_a(\omega-E^{(+)}_k)}{\chi^{(s)}_{p}} \,
         \mem{\chi^{(s)}_{p}}{V_b(E^{(-)}_{q}-\omega)}{\Psi^{(-}_{q}}
       }{(E^{(s)}_{p}-\omega)}
          \\[0.2cm]
\label{e9}
   & = & i\,\sum_p\sum_s 
   \int^{\infty}_{-\infty} \frac{d\omega}{2\pi} 
   \int^{\infty}_{-\infty} d^3\textbf{r} 
     e^{-i(\mathbf{k}-\mathbf{p})\cdot\mathbf{r}}
   A_a(\mathbf{r}; \omega-E^{(+)}_{k})  
   \nonumber \\[0.1cm]
   &   & \hspace*{-0.6cm} \times 		   
   \int^{\infty}_{-\infty} d^3\textbf{r}^{'} \,  
   e^{-i(\mathbf{p}-\mathbf{q})\cdot\mathbf{r}^{'}}
   A_b(\mathbf{r}^{'}; E^{(-)}_{q}-\omega) \,
   \frac{\mem{\textbf{u}^{(+)}_{\sigma_k}}{(1-\beta\alpha_z)}{\textbf{u}^{(s)}_{{\sigma_p}}}
         \mem{\textbf{u}^{(s)}_{\sigma_p}}{(1+\beta\alpha_z)}{
         \textbf{u}^{(-)}_{{\sigma_q}}}}{(E^{(s)}_{p}-\omega)}.
\end{eqnarray}
\end{widetext}
Thus, by combining both integrals  we obtain for the direct free-free pair production amplitude 
the explicit expression:
\begin{eqnarray}\label{e19}
   \mem{\Psi^{(+)}_k}{S_{ab}}{\Psi^{(-)}_q}
   & = & 
   i \sum_s\sum_{\sigma_p}
   \int \frac{d^3\mathbf{p}}{(2\pi)^3} \, 
   \int \frac{d\omega}{2\pi}  \,
   e^{i[\mathbf{p}_\bot-(\frac{\mathbf{k}_\bot+\mathbf{q}_\bot}{2})]\cdot\mathbf{b}} \:
   8\pi^2 Z \gamma^2  
   \frac{\delta(\omega-E^{(+)}_{k} + \beta (k_z-p_z))}
   {(k_z-p_z)^2+\gamma^2(\mathbf{k}_\bot-\mathbf{p}_\bot)^2}
      \nonumber \\[0.1cm]
   &   & \hspace*{0.2cm} \times \,	       
   8\pi^2 Z \gamma^2 
   \frac{\delta(E^{(-)}_{q}-\omega-\beta(p_z-q_z))}{
         (p_z-q_z)^2+\gamma^2(\mathbf{p}_\bot-\mathbf{q}_\bot)^2} \,
   \frac{\mem{\textbf{u}^{(+)}_{\sigma_k}}{(1-\beta\alpha_z)}{\textbf{u}^{(s)}_{\sigma_p}}
         \mem{\textbf{u}^{(s)}_{\sigma_p}}{(1+\beta\alpha_z)}{
              \textbf{u}^{(-)}_{\sigma_q}}
        }{E^{(s)}_{p}-\omega} \, ,
\end{eqnarray}
and where $E^{(s)}_{p}$ is the energy of intermediate state.
In these equations $\mathbf{k}$ represents the the momentum of electron and $\mathbf{q}$ represents the the momentum of positron.
The vector $\mathbf{p}$ describes the momentum of 
the intermediate (electron and positron) states in the field of the ion  and can be
decomposed into its transverse and parallel part, 
$\mathbf{p}=\mathbf{p}_\bot+p_z$, relative to the motion of the ions.
The momenta that is parallel to the heavy ions and the frequency in Eq.~(\ref{d11}) are fixed by momentum conservation;
\begin{subequations}\label{e20}
\begin{eqnarray}\label{e20a}
   p_z=\frac{E^{(-)}_{q}-E^{(+)}_{k}+\beta (k_z+q_z)}{2\beta} \, ,			
\end{eqnarray}
\begin{eqnarray}\label{e20b}
   \omega=\frac{E^{(+)}_{k}+E^{(-)}_{q}-\beta (k_z-q_z)}{2}.	
\end{eqnarray}
\end{subequations}
For the fixed momentum and the spin states, the transition matrix element can be written as;
\begin{widetext}
\begin{eqnarray}\label{e221}
   \mem{\Psi^{(+)}_k}{S_{ab}}{\Psi^{(-)}_q} & = & 
   \frac{i}{2\beta} \, 
   \int \frac{d^2p_\bot}{(2\pi)^2} \, 
   e^{i[\mathbf{p}_\bot-(\frac{\mathbf{k}_\bot+\mathbf{q}_\bot}{2})]\cdot\mathbf{b}} \,
   F(\mathbf{k}_\bot-\mathbf{p}_\bot: \omega_a) \,
   F(\mathbf{p}_\bot-\mathbf{q}_\bot: \omega_b) \,
   \mathcal{T}_{k q} (\mathbf{p}_\bot: +\beta)   \, ,
\end{eqnarray}
\end{widetext}
$\omega_{a}$ and $\omega_{b}$ are the frequencies associated with the fields of ions $a$ and $b$, respectively,

\begin{subequations}\label{e21}
\begin{eqnarray}\label{e21a}
   \omega_{a}=\frac{E^{(-)}_q-E^{(+)}_{k}+\beta (q_z-k_z)}{2} \, ,
\end{eqnarray}
\begin{eqnarray}\label{e21b}
   \omega_{b}=\frac{E^{(-)}_q-E^{(+)}_{k}-\beta (q_z-k_z)}{2} \, ,
\end{eqnarray}
\end{subequations}
the function $F$ is the scalar part of the field from each heavy ion,
\begin{subequations}\label{e22}
\begin{eqnarray}\label{e22a}
   F(\mathbf{k}_\bot-\mathbf{p}_\bot:\omega_a)& = &
   \frac{4\pi Z \gamma^2\beta^2}{
         \left(\omega^{2}_{a} + \gamma^2\beta^2(\mathbf{k}_\bot -
	       \mathbf{p}_\bot)^2\right)} \, ,
\end{eqnarray}
for the frequency $\omega_a$, and as
\begin{eqnarray}\label{e22b}
   F(\mathbf{p}_\bot-\mathbf{q}_\bot:\omega_b) & = &
   \frac{4\pi Ze\gamma^2\beta^2}{
         \left(\omega^{2}_{b} + \gamma^2\beta^2(\mathbf{p}_\bot -
	       \mathbf{q}_\bot)^2\right)} \, ,
\end{eqnarray}
\end{subequations}
for the frequency $\omega_b$, respectively. 
The function $\mathcal{T}$ explicitly depends on the velocity of the heavy ions $\beta$, on the transverse momentum $\mathbf{p}_\bot$, and on the states $k,q$;

\begin{eqnarray}\label{e24}
   &   & \hspace*{-0.75cm}
   \mathcal{T}_{kq}(\mathbf{p}_\bot:+\beta) = 
   \sum_s \sum_{\sigma_p}
   \frac{1}{\left(E^{(s)}_{p} - \left(\frac{E^{(+)}_{k}+E^{(-)}_{q}}{2}\right) 
            +\beta(\frac{k_z-q_z}{2})\right)} 
   \nonumber \\[0.2cm]
   &   & \times
   \mem{\textbf{u}^{(+)}_{\sigma_k}}{(1-\beta\alpha_z)}{\textbf{u}^{(s)}_{\sigma_p}}
   \mem{\textbf{u}^{(s)}_{\sigma_p}}{(1+\beta\alpha_z)}{
        \textbf{u}^{(-)}_{\sigma_q}}  .
\end{eqnarray}
Having the amplitudes for the \textit{direct} and \textit{crossed} diagram, 
we are now prepared to write down the cross section for the generation of a 
free-free pair in collisions of two heavy ions
\begin{eqnarray} \label{ee8}
   \sigma & = & \int d^2b \: \sum_{k>0} \: \sum_{q<0} \:
   \left| \mem{\Psi^{(+)}_k}{S}{\Psi^{(-)}_q} \right|^2 \, ,
\end{eqnarray}
where $S \,=\, S_{ab}+S_{ba}$ denotes the sum of the \textit{direct} and 
\textit{crossed}. Making use of all the simplifications
from above, these cross sections can be expressed as:
\begin{widetext}
\begin{eqnarray}\label{e25}
   \sigma & = & 
   \int d^2b \: \sum_{k>0} \: \sum_{q<0} \:\left|\left\langle 
   \Psi^{(+)}_k\left|S_{ab}\right|\Psi^{(-)}_q\right\rangle+\left\langle     
   \Psi^{(+)}_k\left|S_{ba}\right|\Psi^{(-)}_q\right\rangle\right|^2
   \nonumber \\[0.2cm]	   
   & = &			   
     \sum_{\sigma_k} \sum_{\sigma_q} \int \frac{d^3kd^3qd^2p_\bot}{(2\pi)^8} \,
   \left(\mathcal{A}^{(+)}(k,q;\mathbf{p}_\bot) + 
         \mathcal{A}^{(-)}(k,q;\mathbf{p}_\bot))\right)^2 \,,
\end{eqnarray}
with
\begin{subequations}\label{e26}
\begin{eqnarray}\label{e26a}
   \mathcal{A}^{(+)}(k,q;\mathbf{p}_\bot) & = &
   F(\mathbf{k}_\bot-\mathbf{p}_\bot: \omega_a) \, 
   F(\mathbf{p}_\bot-\mathbf{q}_\bot: \omega_b) \,
   \mathcal{T}_{kq}(\mathbf{p}_\bot:+\beta),				
\end{eqnarray}
and
\begin{eqnarray}\label{e26b}
   \mathcal{A}^{(-)}(k,q;\mathbf{p}_\bot) & = &
   F(\mathbf{k}_\bot-\mathbf{p}_\bot: \omega_b) \,
   F(\mathbf{p}_\bot-\mathbf{q}_\bot: \omega_a) \,
   \mathcal{T}_{kq}(\mathbf{p}_\bot: -\beta) \, .
\end{eqnarray}
\end{subequations}
\end{widetext}
being some proper products of the transition amplitudes and scalar parts 
of the fields as associated with ions $a$ and $b$. 
 
\section{Calculations and Tables}

Table~\ref{t1} displays the cross sections of electromagnetic productions of electron, muon and tauon pairs from the ultra-relativistic gold-gold collisions at RHIC energies and lead-lead collisions at LHC energies with no form factor and with Wood-Saxon form factor distribution. In this calculation, we simply used the analytical expression of the Fourier transformation of this form factor. We have also include the electric distribution function of the proton (Eq. 9) in our calculation, and we have seen that it has negligible effect for the production of any lepton pairs.

When we compare electron pair production cross sections with and without form factor, it is clearly seen that the difference among the results are negligibly small. Since the Compton wavelength of the electron is much larger than the radius of the colliding heavy ions, 
form factors of the proton and nucleus almost have no effect for the production of electron-positron pairs.

However, when we compare muon and tauon pair production cross sections with and without form factors, it is clearly seen that the differences are large. The cross section results for muon and tauon pair production with form factors are much smaller than without form factors. Since, the Compton wavelengths of muons and tauons are smaller than the radius of the colliding heavy ions, the effects of the form factors become dominant for pair production. At RHIC and LHC energies, muon pair production is reduced by about 3 and 2 factors, respectively. On the other hand, at RHIC and LHC energies, tauon pair production is reduced by about 100 and 5 factors, respectively.

\begin{figure}[h]
  \centering
  \subfloat[]{\includegraphics[width=5.5cm,height=4.5cm]{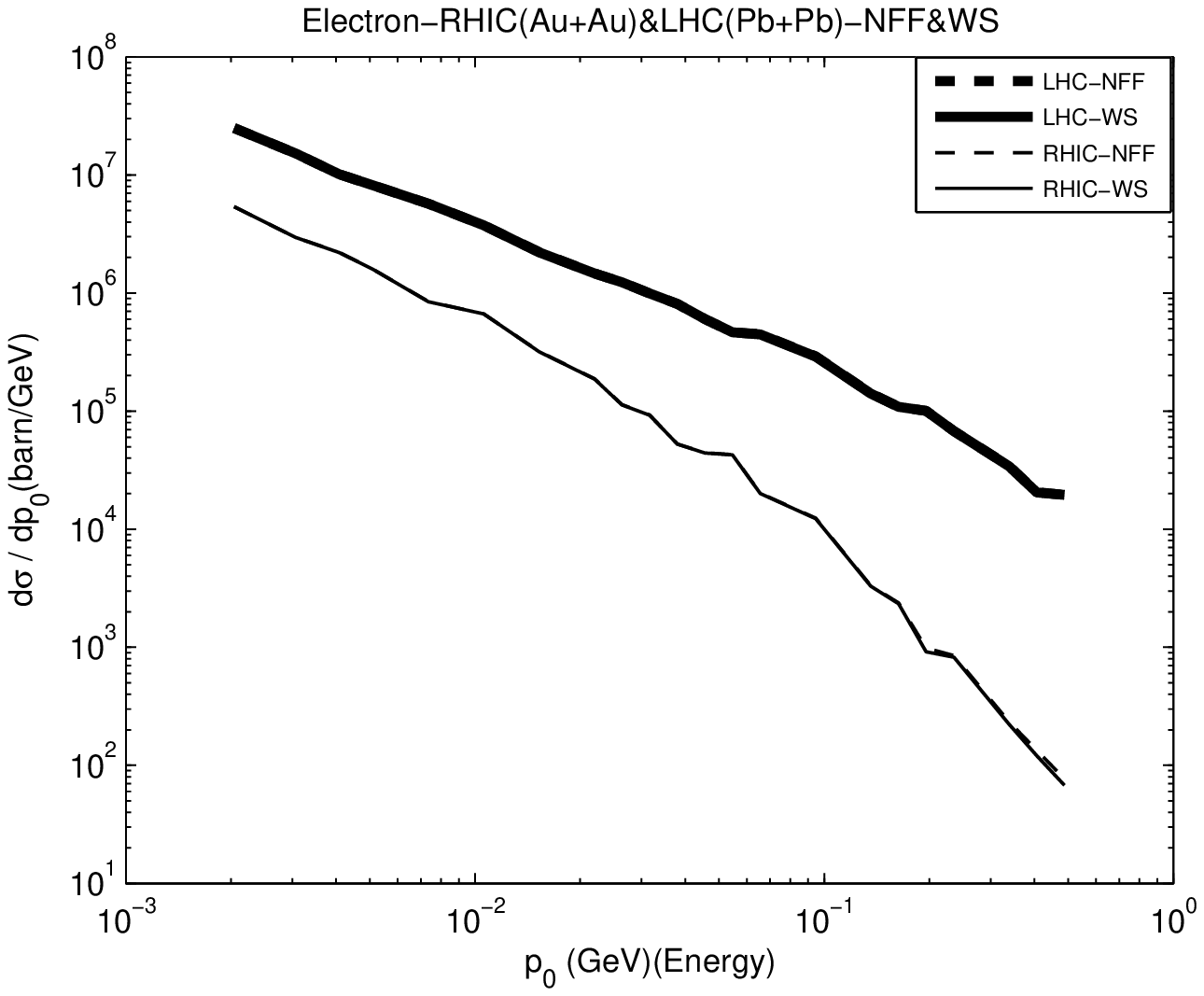}}                
  \subfloat[]{\includegraphics[width=5.5cm,height=4.5cm]{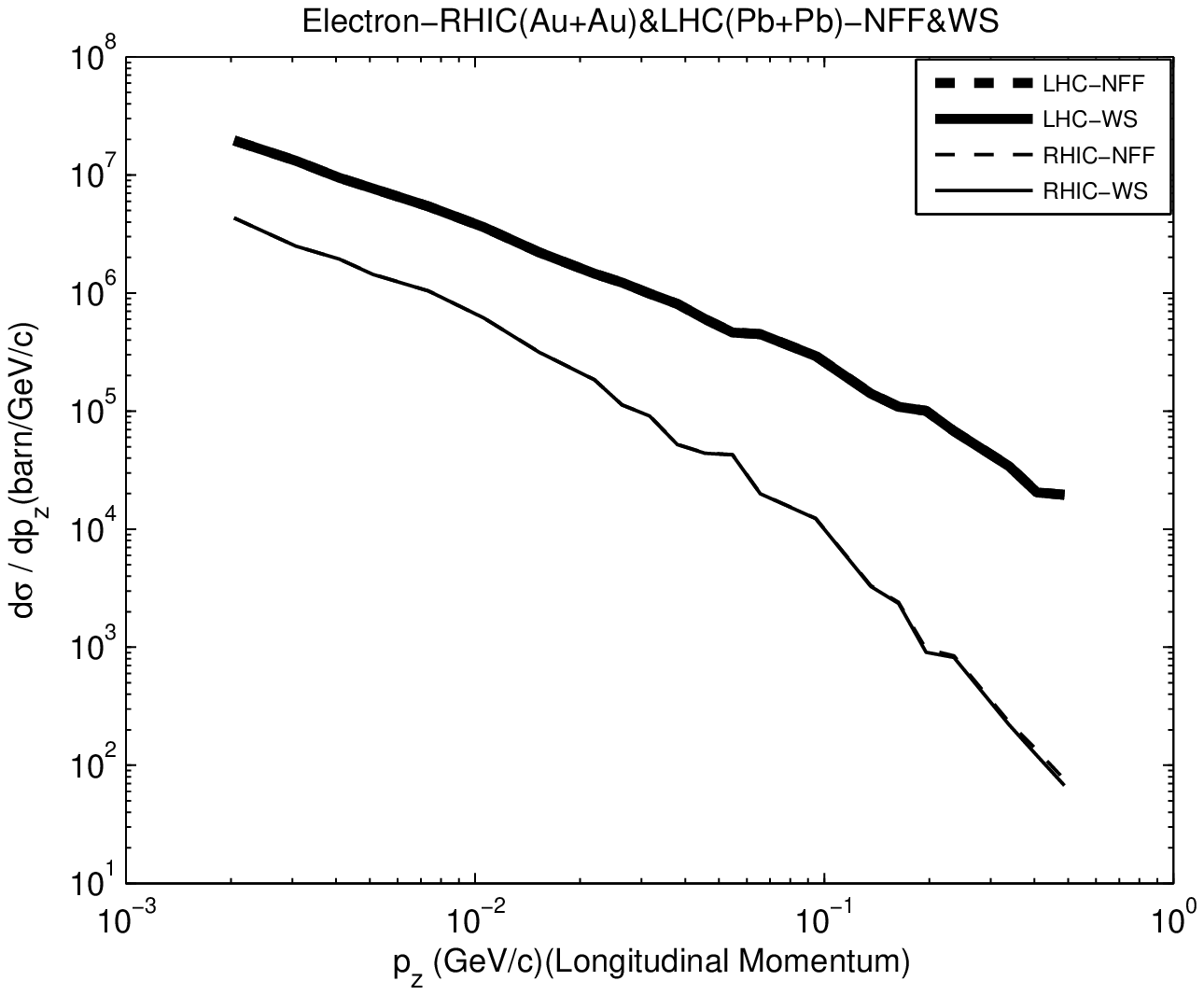}}\\
  \subfloat[]{\includegraphics[width=5.5cm,height=4.5cm]{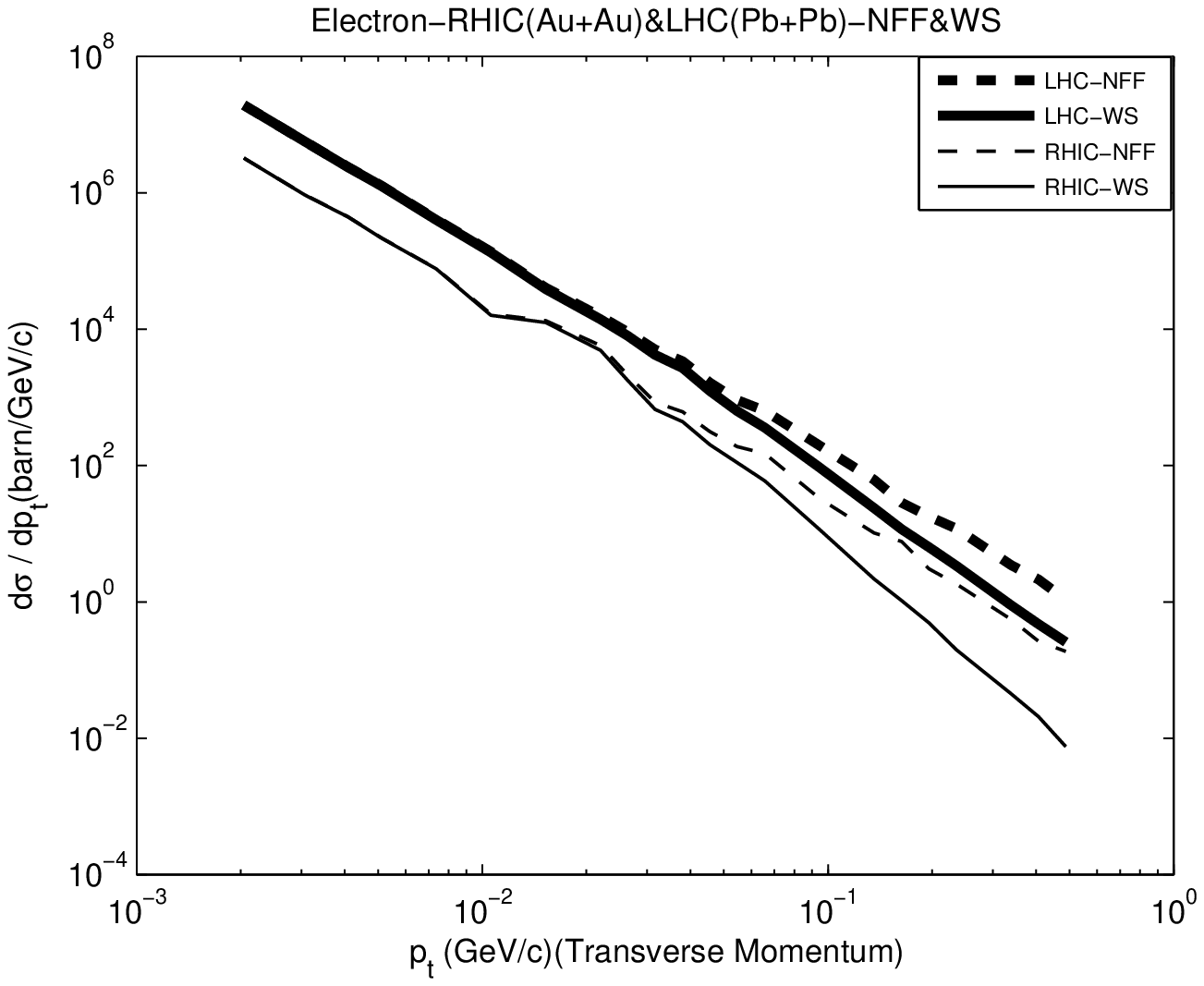}}                
  \subfloat[]{\includegraphics[width=5.5cm,height=4.5cm]{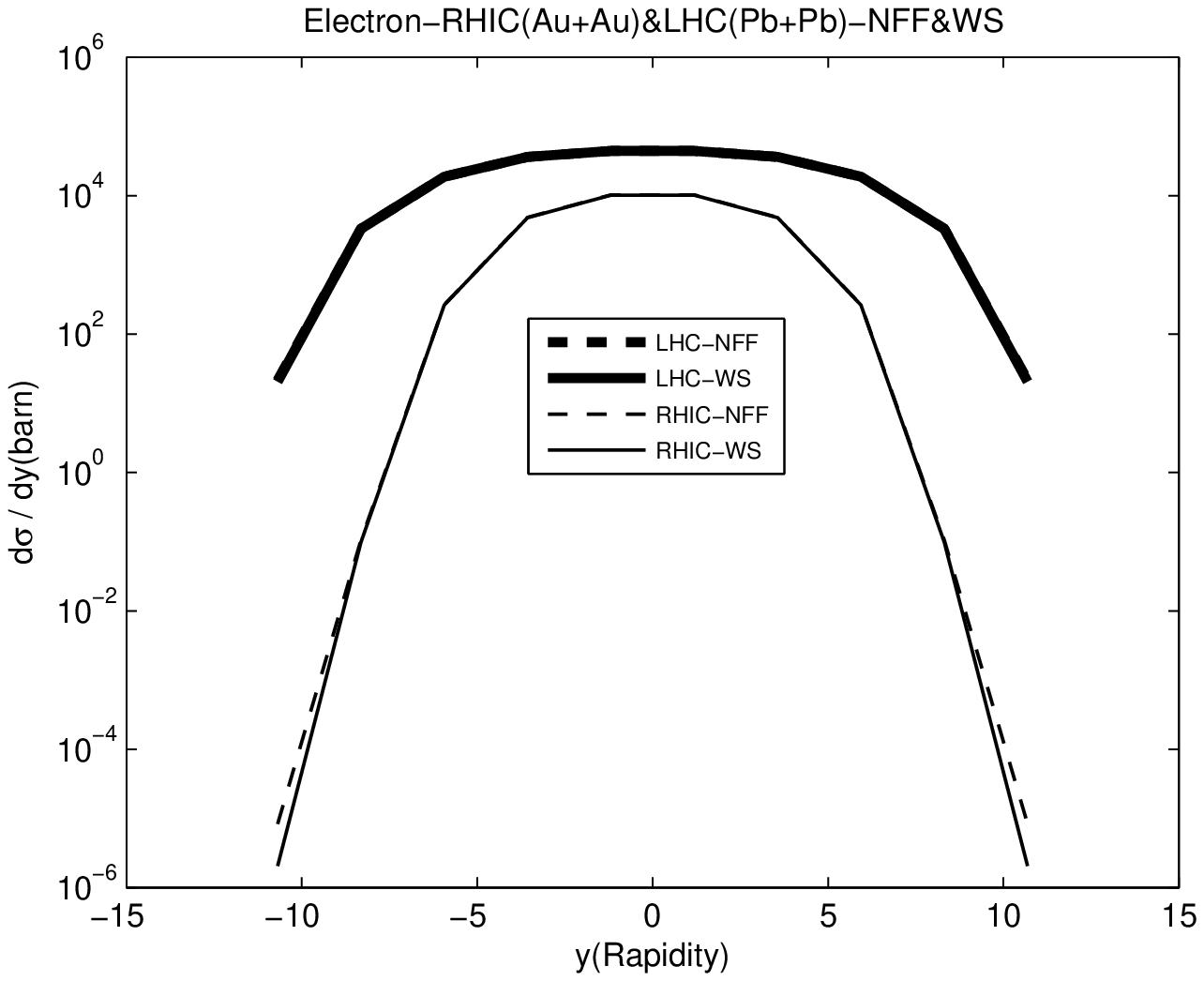}}
  \caption{The differential cross section as function of  a) energy ($p_{0}$), b) longitudinal momentum ($p_{z}$), c) transverse momentum ($p_{\bot}$) and d) rapidity ($y$) of the produced electron at RHIC and LHC energies with and without form factors.}
\label{f231}
\end{figure}

\begin{figure}[h]
  \centering
  \subfloat[]{\includegraphics[width=5.5cm,height=4.5cm]{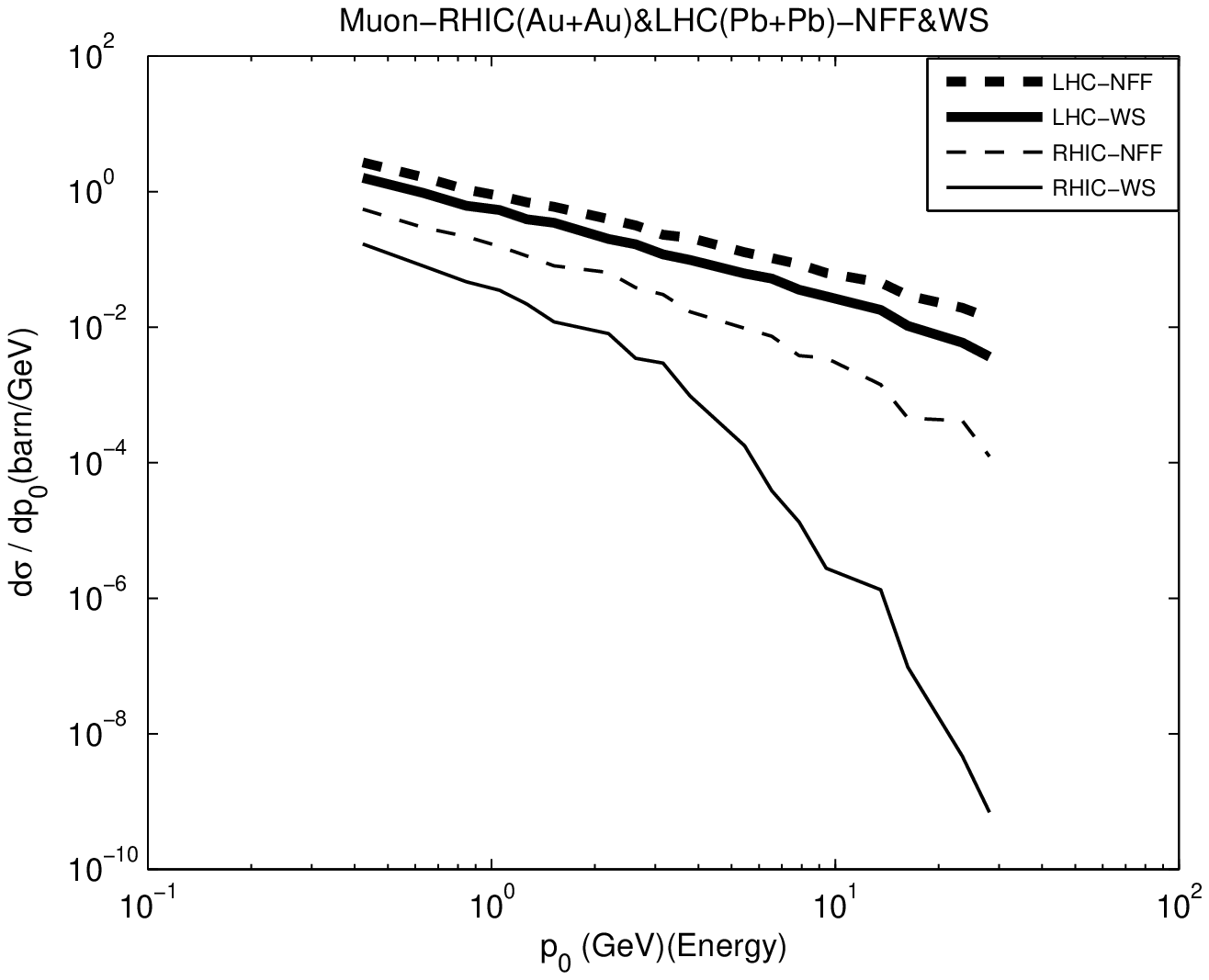}}                
  \subfloat[]{\includegraphics[width=5.5cm,height=4.5cm]{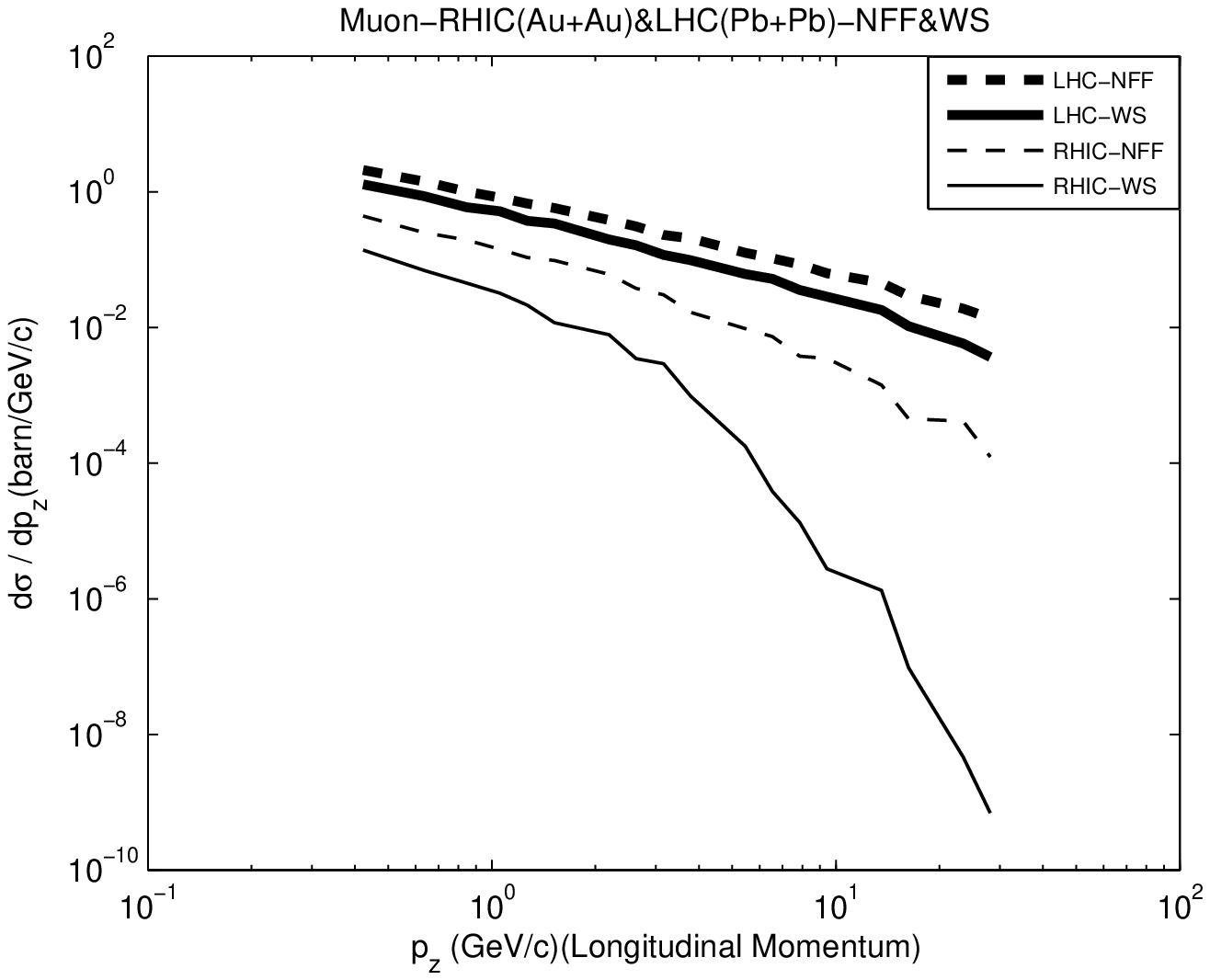}}\\
  \subfloat[]{\includegraphics[width=5.5cm,height=4.5cm]{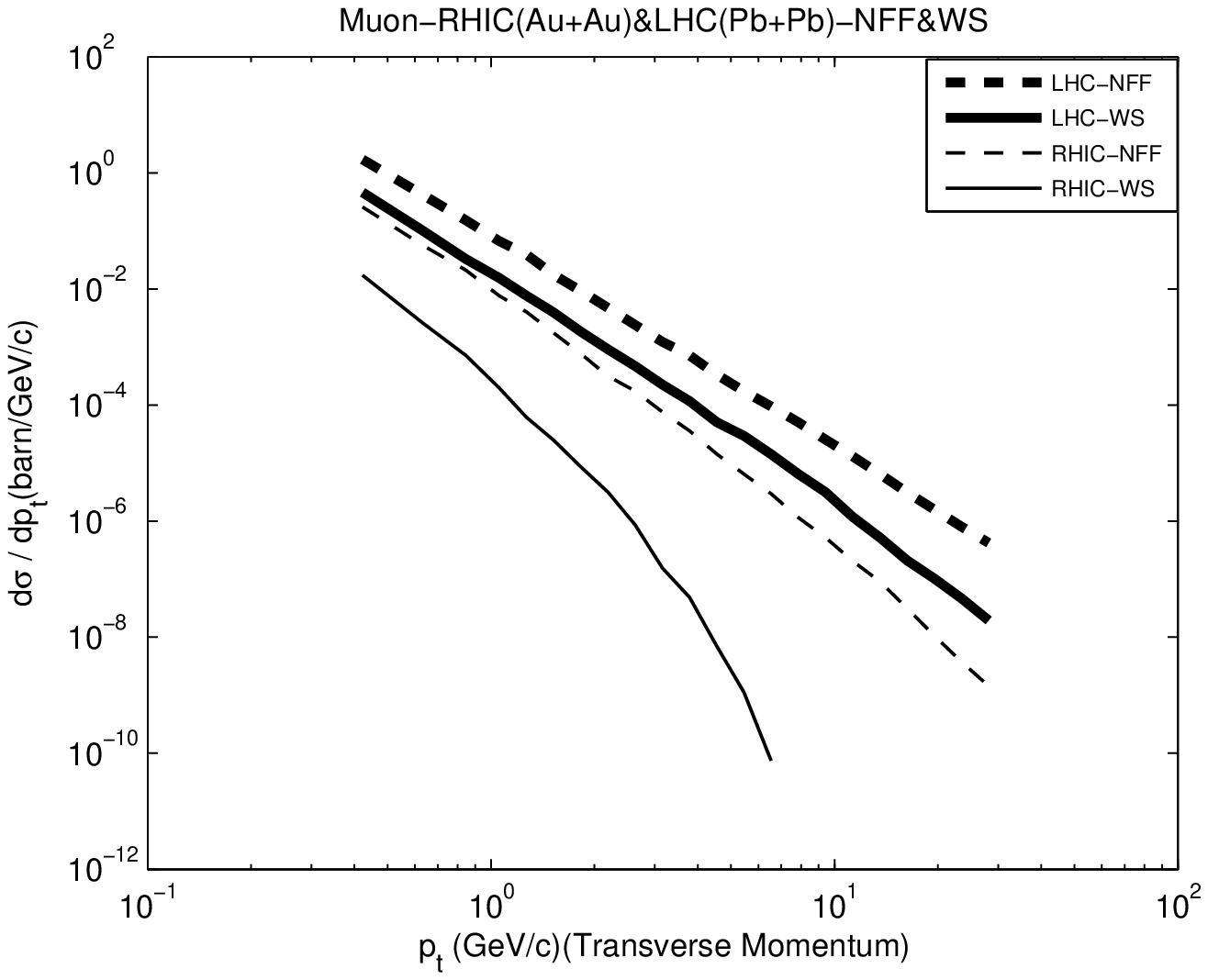}}                
  \subfloat[]{\includegraphics[width=5.5cm,height=4.5cm]{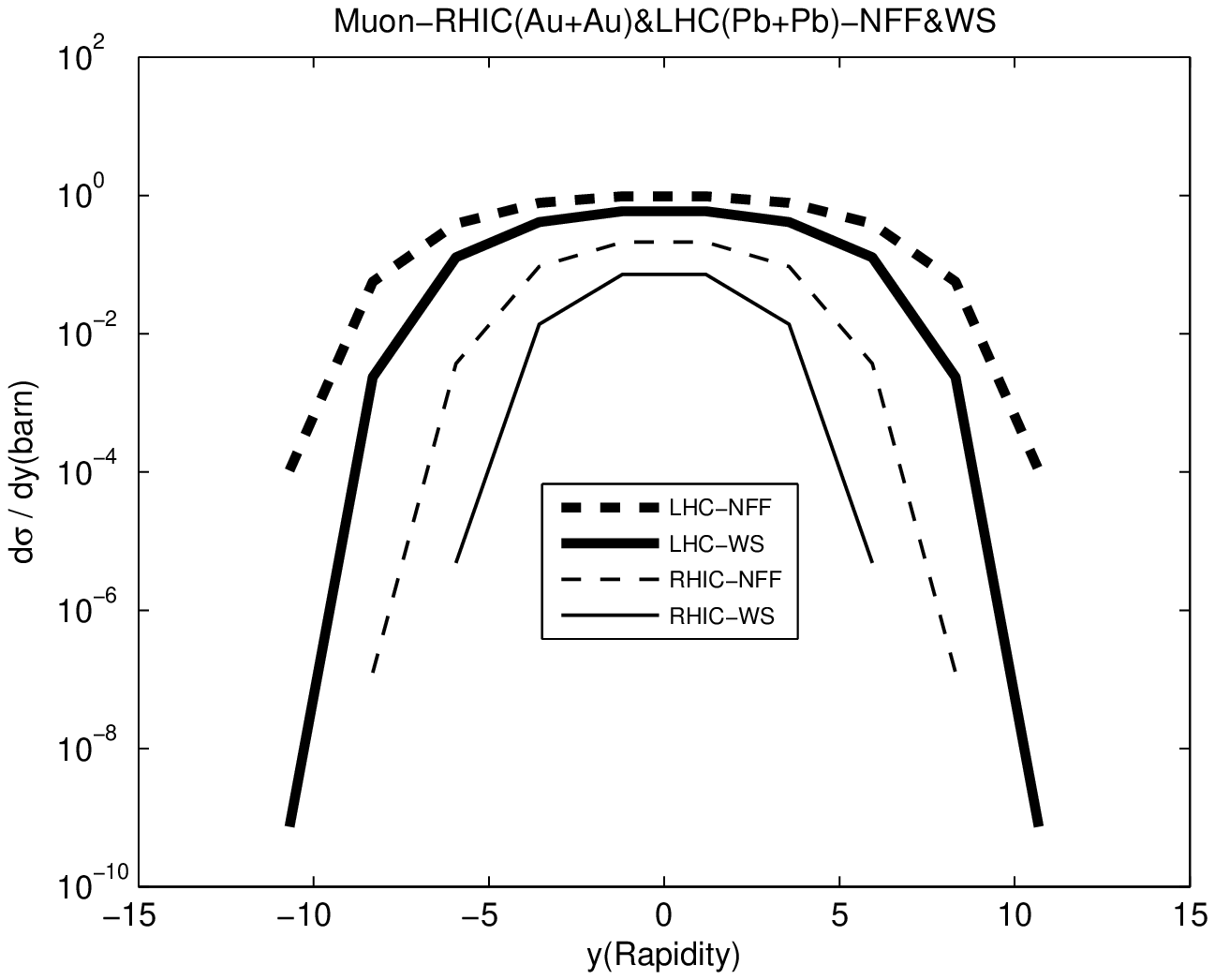}}\\
  \caption{The differential cross section as function of a) energy ($p_{0}$), b) longitudinal momentum ($p_{z}$), c) transverse momentum ($p_{\bot}$) and d) rapidity ($y$) of the produced muon at RHIC and LHC energies with and without form factors.}
\label{f232}
\end{figure}

\begin{figure}[h]
  \centering
  \subfloat[]{\includegraphics[width=5.5cm,height=4.5cm]{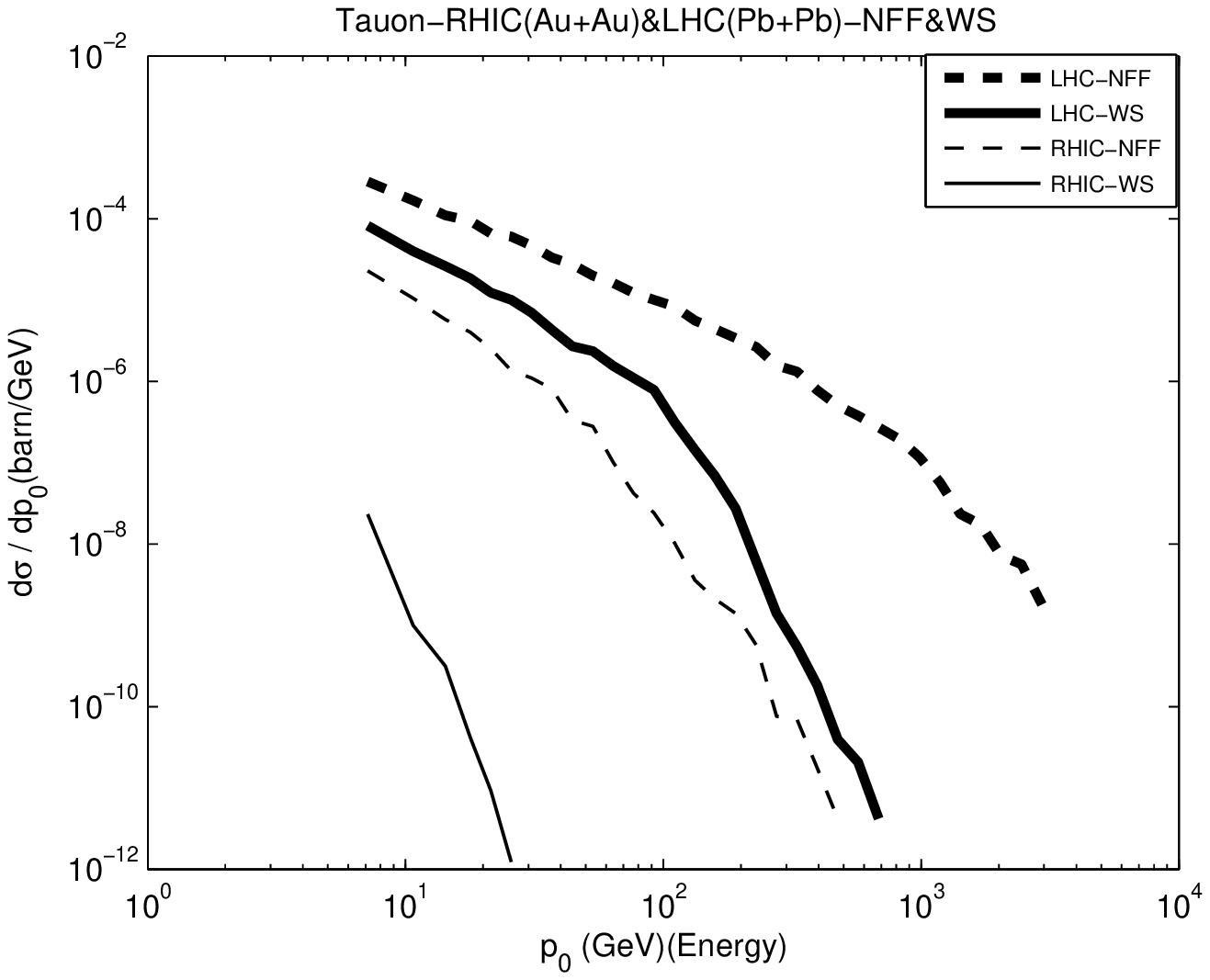}}                
  \subfloat[]{\includegraphics[width=5.5cm,height=4.5cm]{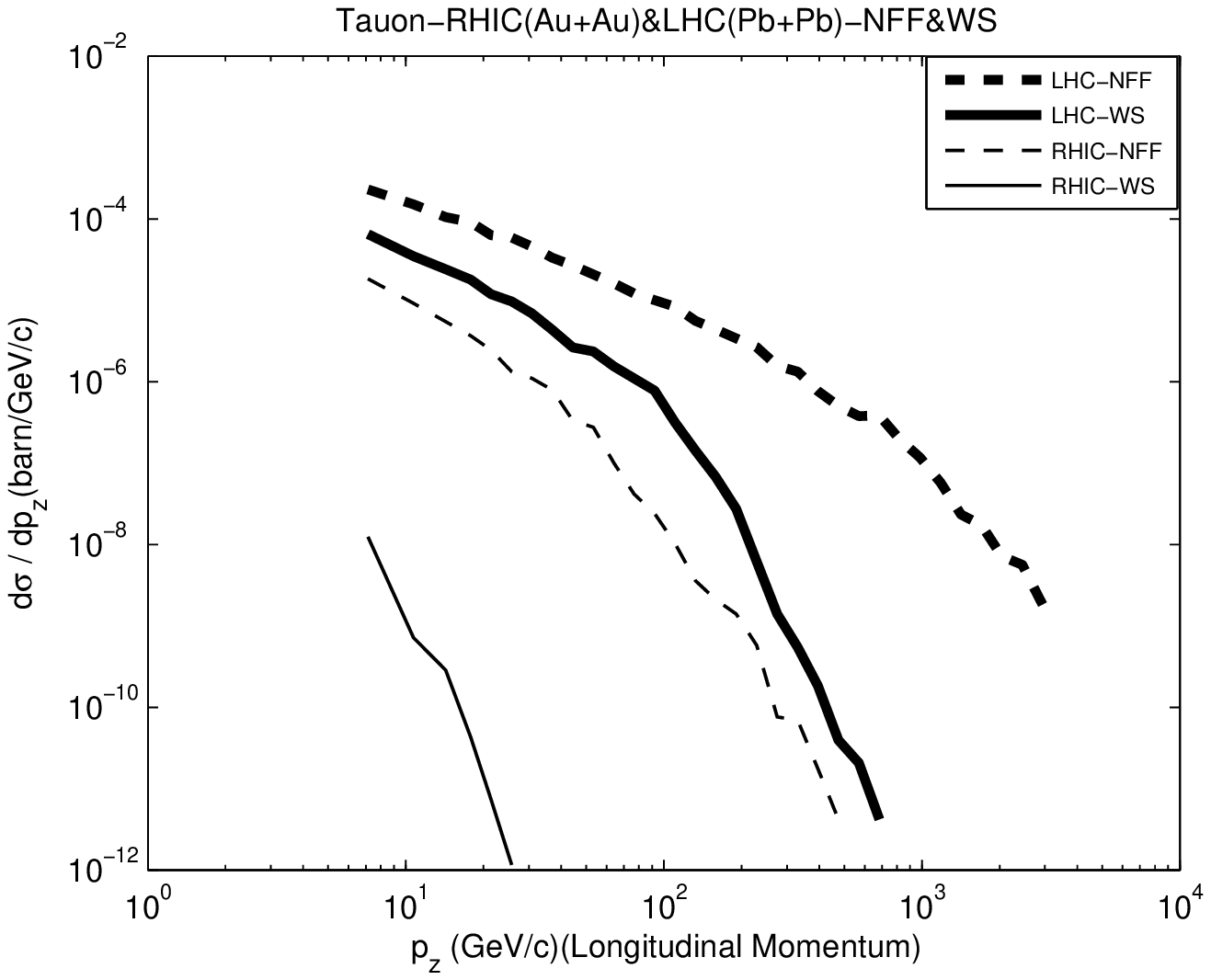}}\\
  \subfloat[]{\includegraphics[width=5.5cm,height=4.5cm]{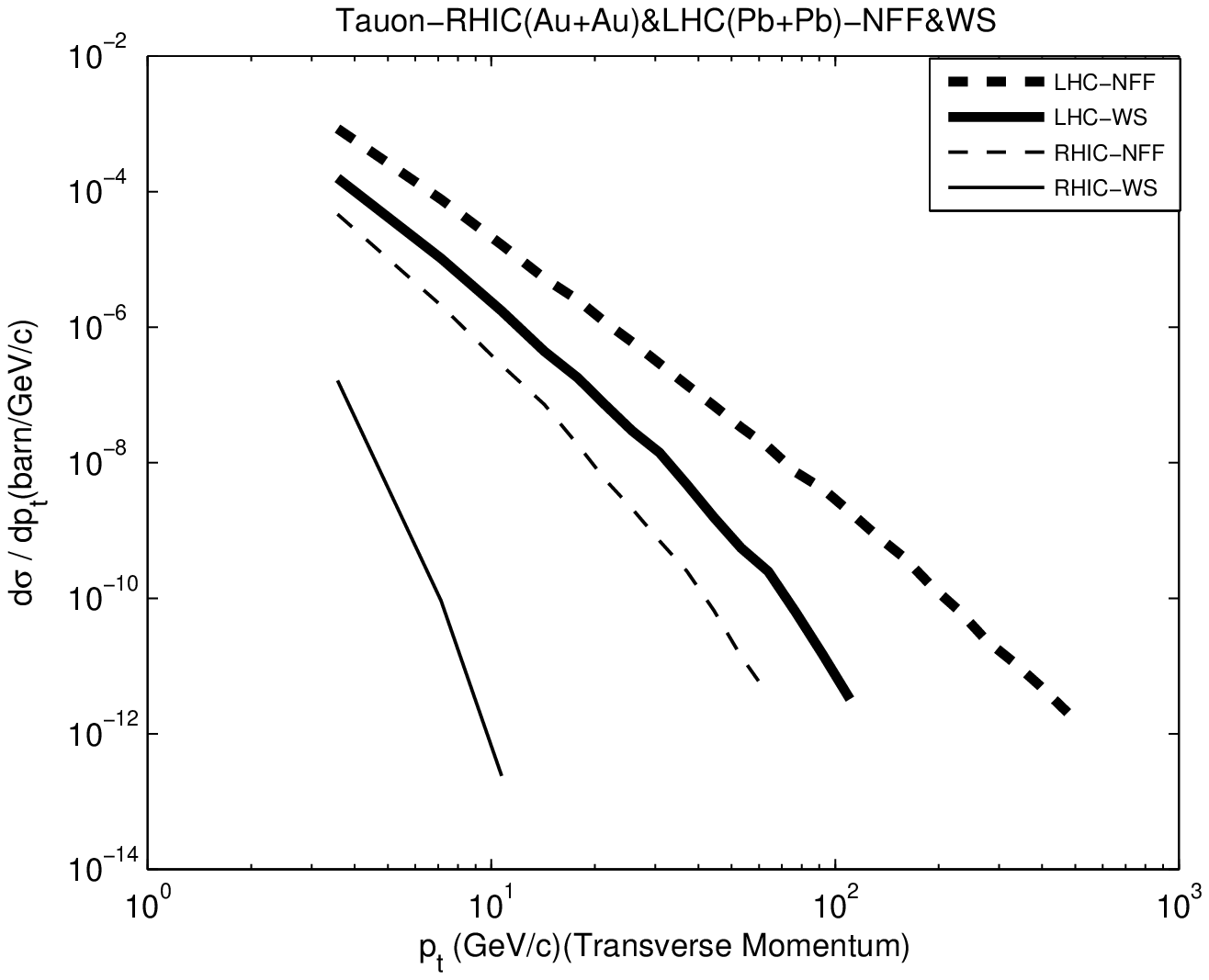}}   
  \subfloat[]{\includegraphics[width=5.5cm,height=4.5cm]{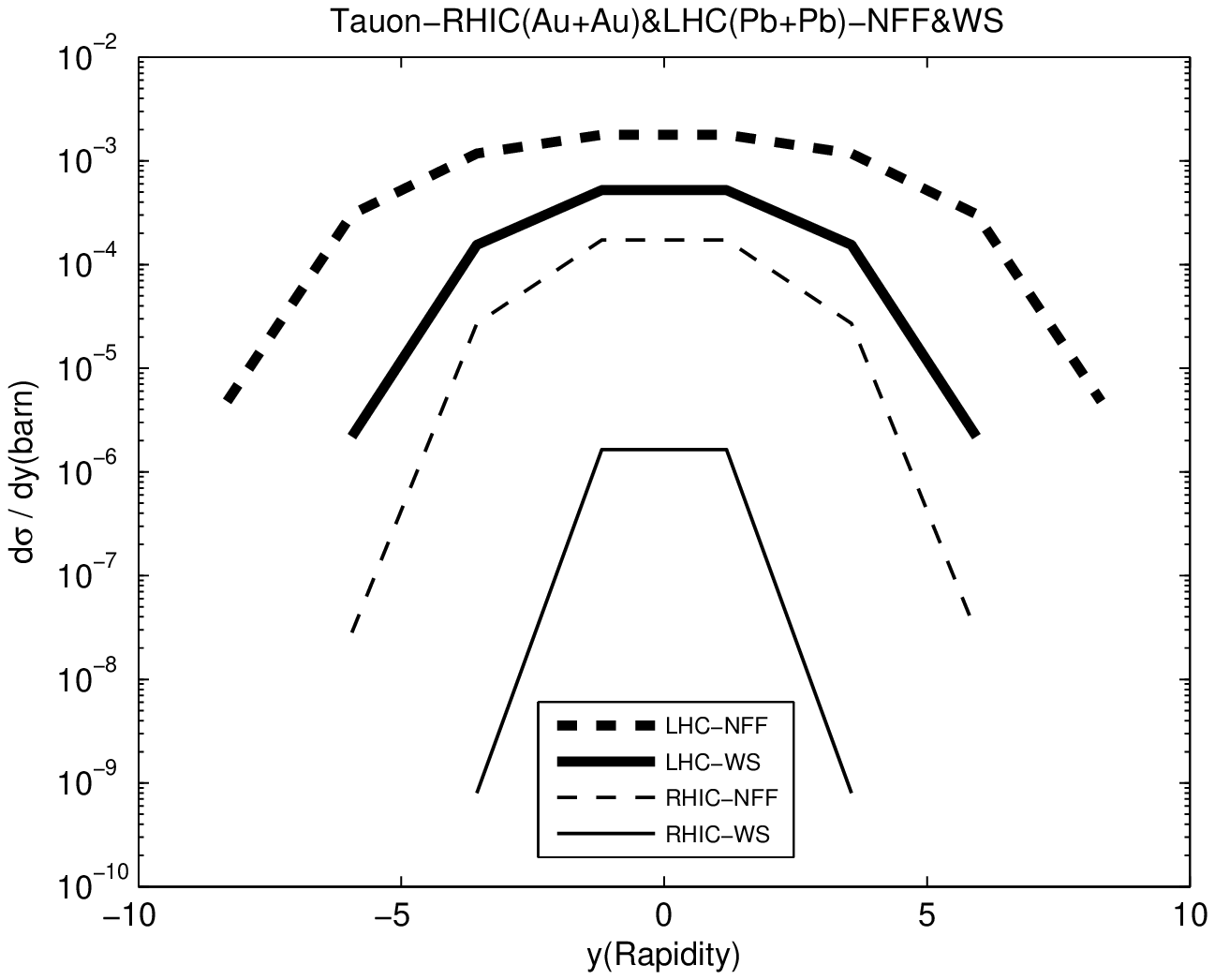}}\\
  \caption{The differential cross section as function of a) energy ($p_{0}$), b) longitudinal momentum ($p_{z}$), c) transverse momentum ($p_{\bot}$) and d) rapidity ($y$) of the produced tauon  at RHIC and LHC energies with and without form factors.}
\label{f24}
\end{figure}

\begin{table}[t]
\caption{The cross sections of electromagnetic productions of electron, muon and tauon pair productions from the ultra-relativistic gold-gold collisions at RHIC energies and lead-lead collisions at LHC energies.  $f^{NFF}$ means form factor is not included and $f^{WS}$ means that form factor is included in the calculations.}\label{t1}\
\begin{center}
\begin{tabular}{ll l l}
\hline\hline
\rule{0pt}{1em}$ $ &  &  & \\
\rule{0pt}{1em}$ $ & & \multicolumn{1}{c}{$\sigma^{f^{NFF}}$(barn)}\,\,\,\,\,\,\,\,\, & \multicolumn{1}{c}{$\sigma^{f^{WS}}$(barn)}\\
\hline \\
\multicolumn{1}{l}{} $e^{-}e^{+}$ & RHIC & $3.63\times10^{4}$ & $3.62\times10^{4}$ \\ 
\multicolumn{1}{l}{}                      & LHC  & $2.44\times10^{5}$ & $2.43\times10^{5}$ \\ \\
\multicolumn{1}{l}{} $\mu^{-}\mu^{+}$ & RHIC &  $7.3\times10^{-1}$ & $2.03\times10^{-1}$ \\ 
\multicolumn{1}{l}{}                      & LHC  & $5.2$ & $2.7$ \\ \\
\multicolumn{1}{l}{} $\tau^{-}\tau^{+}$ & RHIC & $4.7\times10^{-4}$ & $3.9\times10^{-6}$ \\ 
\multicolumn{1}{l}{}                      & LHC  & $7.7\times10^{-3}$ & $1.6\times10^{-3}$  \\ \\ \hline\hline
\end{tabular}
\end{center}
\end{table}
\begin{table}[t]
	\centering
	\caption{The percentage of cross section contributions to the total cross section between the impact parameter regions $0<b<15-20-25-30 fm$ and $15-20-25-30 fm<b<\infty$. The cross section calculations are taken from the Table I with form factor.}\label{t2}\
	\begin{tabular}{c c c c c c c }
		\hline\hline \\
		& \multicolumn{2}{c}{$e^{-}e^{+}$} & \multicolumn{2}{c}{$\mu^{-}\mu^{+}$} & \multicolumn{2}{c}{$\tau^{-}\tau^{+}$} \\ 
		&$b<15 fm$& \,\,\,\,\,$b> 15 fm$ & \,\,\,\,\, $b<15 fm$ & \,\,\,\,\,$b> 15fm$ & \,\,\,\,\,$b<15 fm$ &\,\,\,\,\,$b>15fm$ \\ \hline \\
		RHIC&  0.04$\%$      &    99.96$\%$       &  13.2$\%$         &   86.8$\%$        &    49.8$\%$       &     50.2$\%$      \\ 
		LHC&   0.04$\%$    &      99.96$\%$     &    3.65$\%$       &   96.35$\%$        &    25$\%$       &      75$\%$ \\ \\ 
		&$b<20 fm$& \,\,\,\,\,$b> 20 fm$ & \,\,\,\,\, $b<20 fm$ & \,\,\,\,\,$b> 20fm$ & \,\,\,\,\,$b<20 fm$ &\,\,\,\,\,$b>20fm$ \\ \hline \\
		RHIC&  0.07$\%$      &    99.93$\%$       &  20.44$\%$         &   79.56$\%$        &    60.11$\%$       &     39.89$\%$      \\ 
		LHC&   0.07$\%$    &      99.93$\%$   &      6.24$\%$       &   93.76$\%$        &    35.23$\%$       &      64.77$\%$ \\ \\ 
	  &$b<25 fm$& \,\,\,\,\,$b> 25 fm$ & \,\,\,\,\, $b<25 fm$ & \,\,\,\,\,$b> 25fm$ & \,\,\,\,\,$b<25 fm$ &\,\,\,\,\,$b>25fm$ \\ \hline \\
		RHIC&  0.11$\%$      &    99.89$\%$       &  27.6$\%$         &   72.4$\%$        &    67.13$\%$       &     32.87$\%$      \\ 
		LHC&   0.11$\%$    &      99.89$\%$   &     9.27$\%$       &   90.73$\%$        &    43.77$\%$       &      56.23$\%$ \\ \\
	  &$b<30 fm$& \,\,\,\,\,$b> 30 fm$ & \,\,\,\,\, $b<30 fm$ & \,\,\,\,\,$b> 30fm$ & \,\,\,\,\,$b<30 fm$ &\,\,\,\,\,$b>30fm$ \\ \hline \\
		RHIC&  0.16$\%$      &    99.84$\%$       &  34.15$\%$         &   65.85$\%$        &    72.15$\%$       &     27.85$\%$      \\ 
		LHC&   0.16$\%$    &      99.84$\%$   &    12.6$\%$       &   87.4$\%$        &    50.7$\%$       &      49.3$\%$ \\ \\ \hline\hline
	\end{tabular}
\end{table}
\begin{table}[t]
	\caption{The ratio of the cross sections ($\Delta \sigma /\sigma$) of the produced muon and tauon pairs. At the first column, the parameter $R$ is kept constant and the parameter  $a$ is changed between $ 0.4 fm\leq a \leq 0.6 fm$ ($\Delta a = 0.2 fm$). 
	At the second column, the parameter $a=0.55 fm$ is kept constant and the radius $R$ is changed between $6.1 fm \leq R \leq 7.0 fm $.}\label{t3}\
	\begin{center}
		\begin{tabular}{ll c c}
			\hline\hline
			\rule{0pt}{1em}$ $ &  &  & \\
			\rule{0pt}{1em}$ $ & & \multicolumn{1}{c}{$R_{Au}=6.38 fm$,$R_{Pb}=6.62 fm$\,\,$\Delta a = 0.2 fm$}\,\,\,\,\,\,\,\,\, & \multicolumn{1}{c}{$a=0.55 fm$,\,\,$\Delta R = 0.9 fm$}\\
			\rule{0pt}{1em}$ $ & & \multicolumn{1}{c}{$\Delta \sigma /\sigma$}\,\,\,\,\,\,\,\,\, & \multicolumn{1}{c}{$\Delta \sigma /\sigma$}\\
			\hline \\
			\multicolumn{1}{l}{} $\mu^{-}\mu^{+}$ & RHIC &  4 $\%$ & 11 $\%$ \\ 
			\multicolumn{1}{l}{}                      & LHC  & 2 $\%$ & 5 $\%$ \\ \\
			\multicolumn{1}{l}{} $\tau^{-}\tau^{+}$ & RHIC & 14 $\%$ & 35 $\%$ \\ 
			\multicolumn{1}{l}{}                      & LHC  & 4 $\%$ & 10 $\%$  \\ \\ \hline\hline
		\end{tabular}
	\end{center}
\end{table}
Figs.~\ref{f231}, \ref{f232} and \ref{f24} displays the differential cross sections
as a function of the energy $p_{0}$, the longitudinal momentum $p_{z}$, the transverse momentum $p_{\perp}$ and rapidity $y$ of the produced electron, muon
and tauon for with and without form factors at RHIC and LHC energies. When we compare these figures, in Fig.\ref{f231} for the electron pairs, it is clearly seen that form factor has almost no effect to the cross section.
With and without form factors, differential cross sections have the same behavior and almost identical values. The differential cross sections as a function of transverse momentum decreases rapidly compare to the longitudinal momentum and energy. Differential cross sections of as a function longitudinal momentum and energy have almost identical values. This also shows that produced leptons carry energy in the longitudinal direction where the heavy-ions move.

On the other hand, In Fig.\ref{f232} and \ref{f24} the same differential cross sections of the produced muon and tauon pairs at RHIC and LHC energies with and without form factors are shown. When we compare these figures, it is clear that the longitudinal momentum of the produced muon and tauon are much higher than the transverse momentum. Therefore, the energy $E=\sqrt{p^{2}_{t}+p^{2}_{z}+1} $ are carried by the longitudinal momentum of the produced heavy leptons. It is also clear that, the differential cross sections as a function of energy and longitudinal momentum are very close to each other at RHIC and LHC energies. The rapidity distribution of muon pair production, form factor reduce the cross sections by about ten times at the RHIC energies, and reduce the cross section by about five times at the LHC energies. For the tauon pair production, the reduction of the cross section is about two order of magnitude for the RHIC energies, and is about one order of magnitude for the LHC energies. All these calculations show that distributions of the heavy leptons are mainly confined to the central rapidity region.

\begin{figure}[h]
  \centering
  \subfloat[]{\includegraphics[width=5.5cm,height=4.5cm]{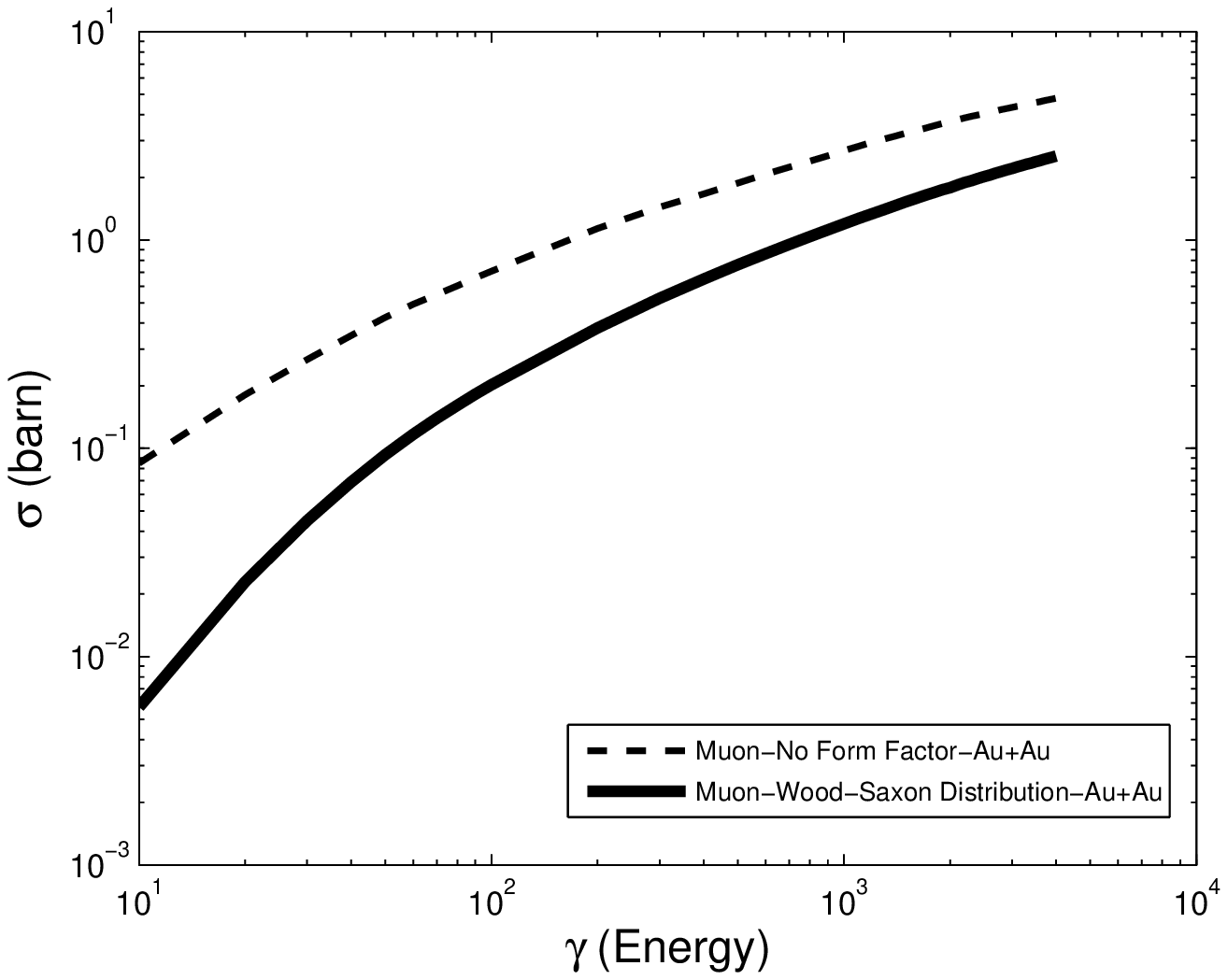}}                
  \subfloat[]{\includegraphics[width=5.5cm,height=4.5cm]{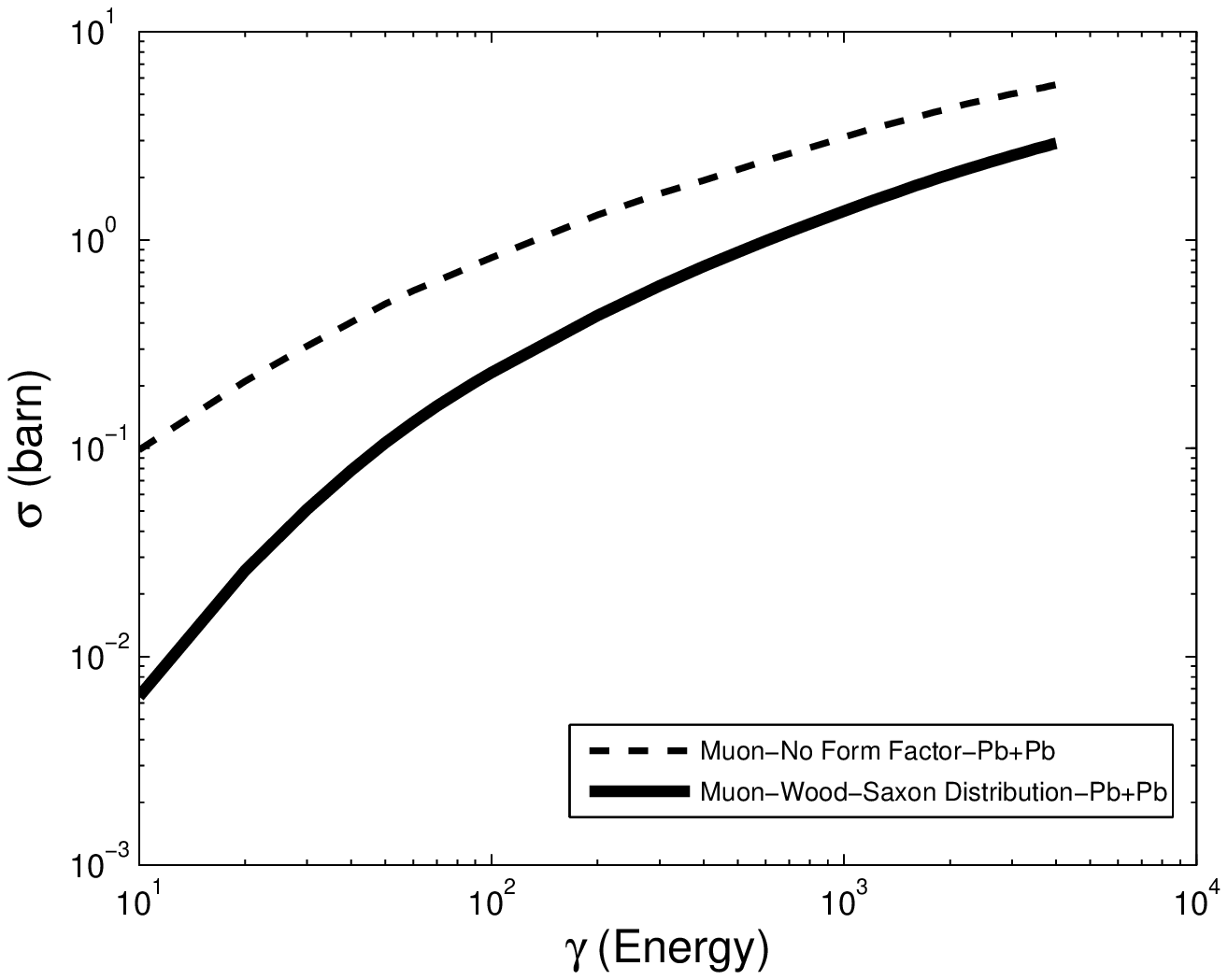}}\\
  \subfloat[]{\includegraphics[width=5.5cm,height=4.5cm]{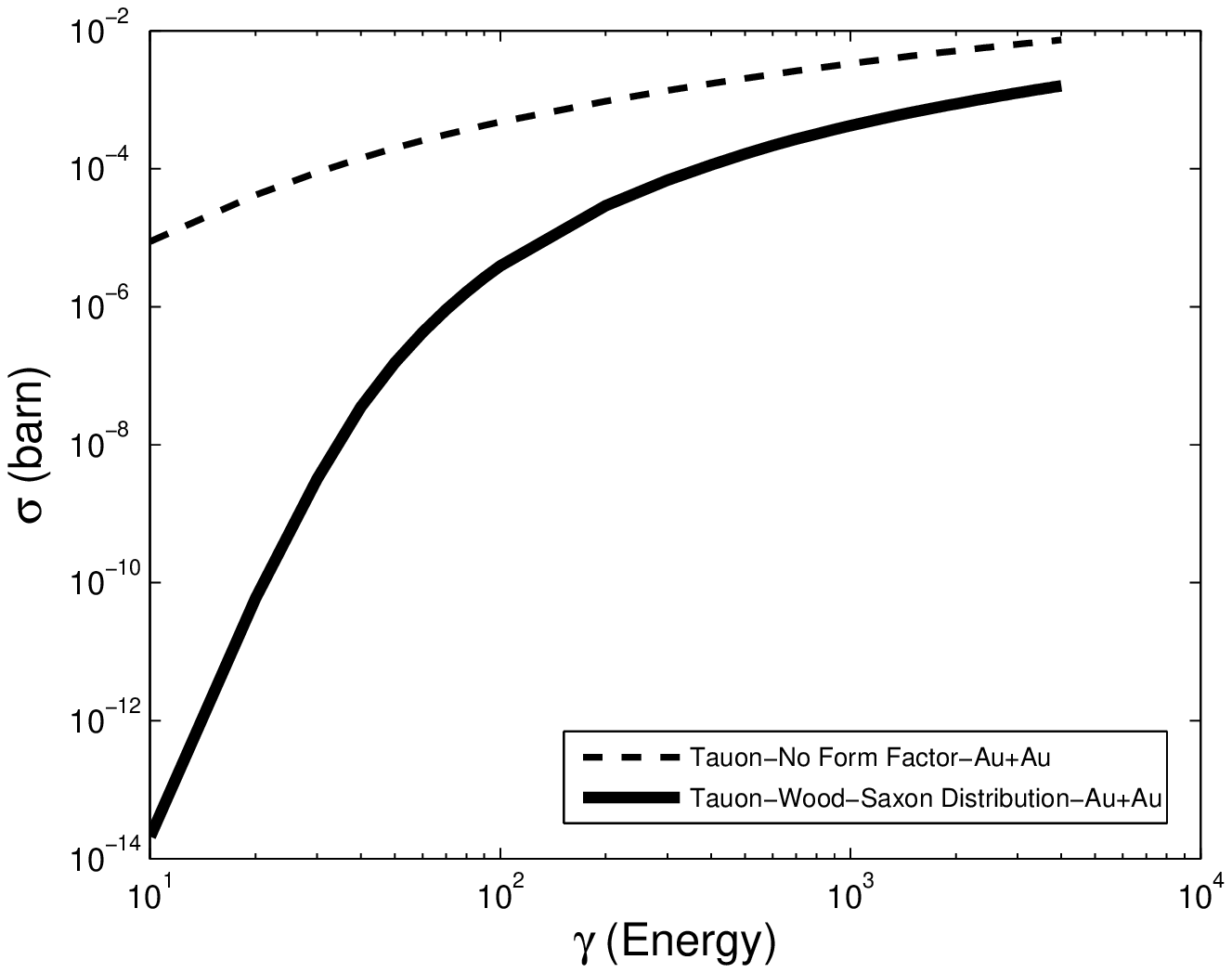}}               
  \subfloat[]{\includegraphics[width=5.5cm,height=4.5cm]{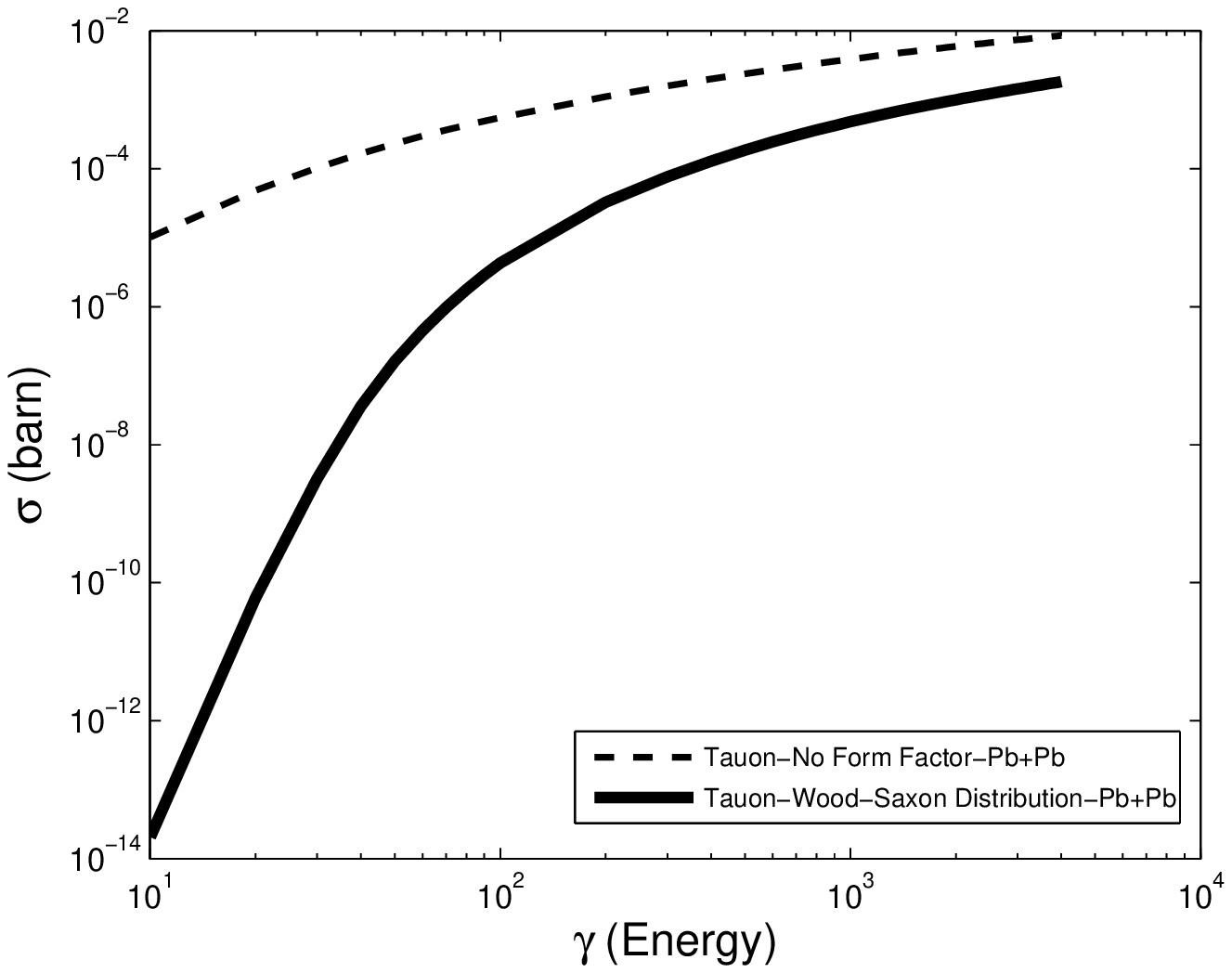}}
  \caption{Muon and tauon cross sections for $Au+Au$ and $Pb+Pb$ collisions as function of the $\gamma$ with and without form factors.}
\label{f25}
\end{figure}

Fig.~\ref{f25} displays muon and tauon total cross sections for $Au+Au$ and $Pb+Pb$ collisions as a function of the Lorentz factor $\gamma$ of the beams for the Wood-Saxon form factors. We also compare this results with the no form factor. For the muon pair production, the total cross sections are reduced about 10 factors for the low energies, about 3 factors for the RHIC energies, and about 2 factors for the LHC energies. For the tauon pair production,
the total cross sections are reduced about $10^{9} - 10^{6}$ factors for the low energies, about 100 factors for the RHIC energies, and finally about 5 factors for the LHC energies.
\begin{figure}[h]
  \centering
  \subfloat[]{\includegraphics[width=5.5cm,height=4.5cm]{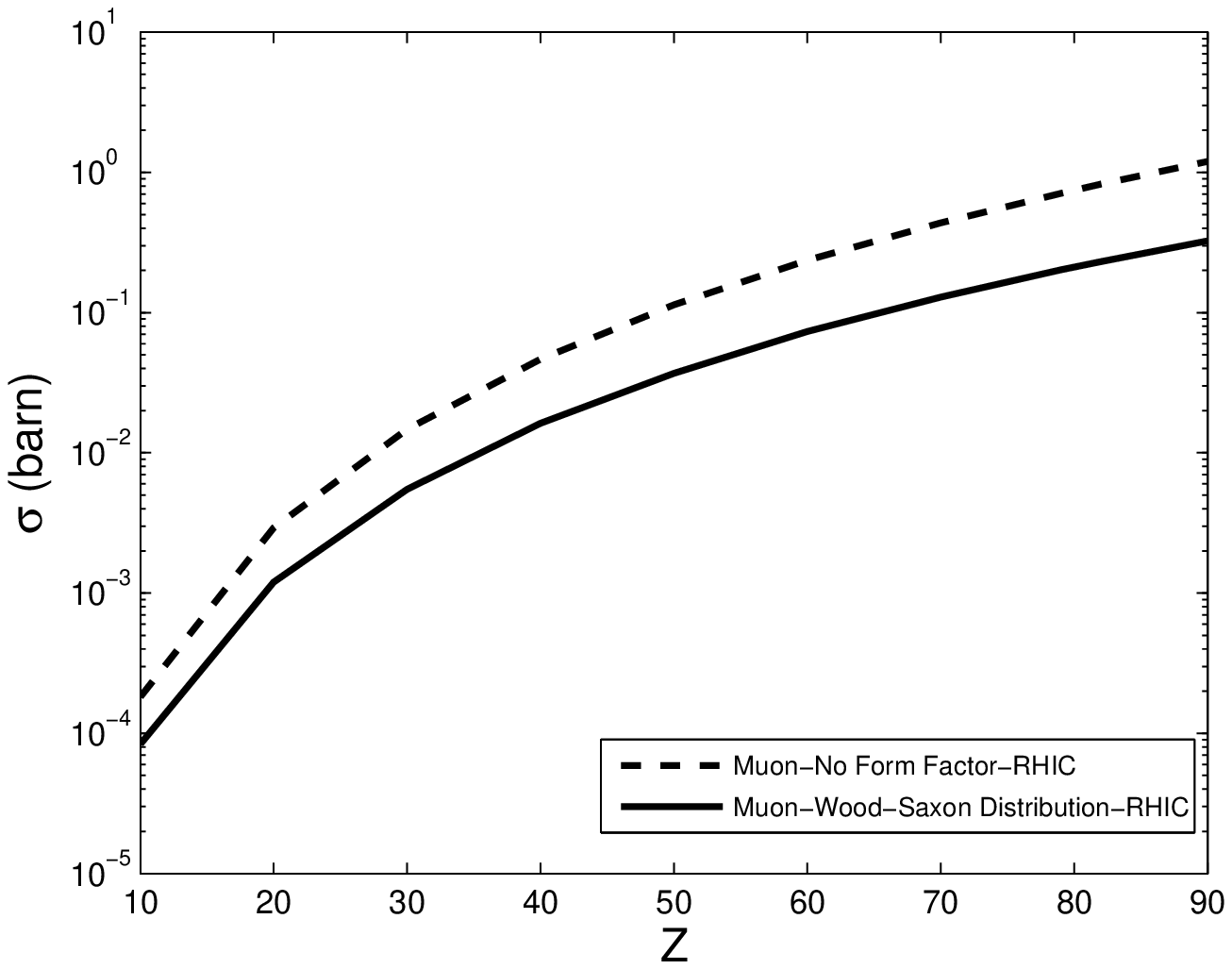}}                
  \subfloat[]{\includegraphics[width=5.5cm,height=4.5cm]{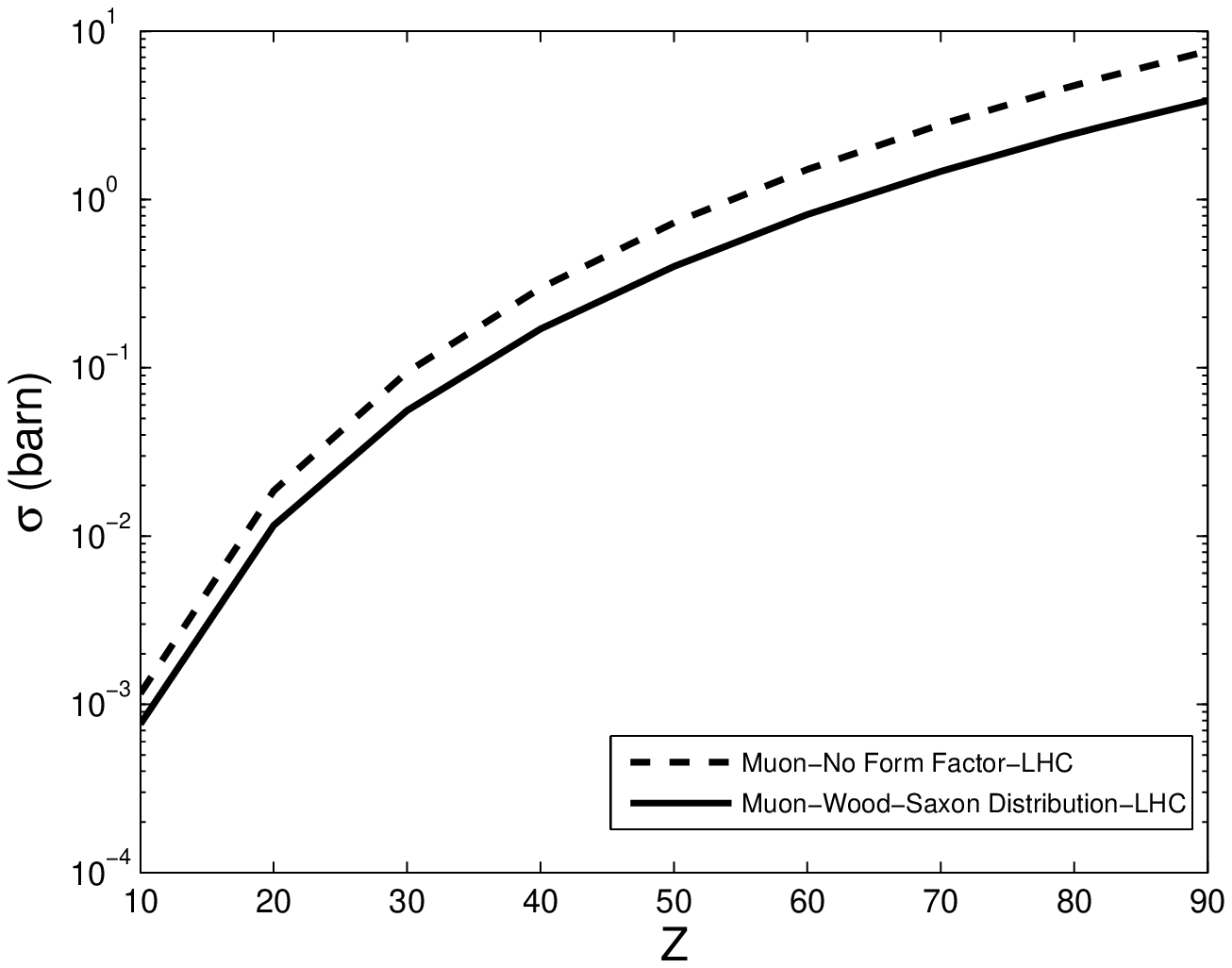}}\\
  \subfloat[]{\includegraphics[width=5.5cm,height=4.5cm]{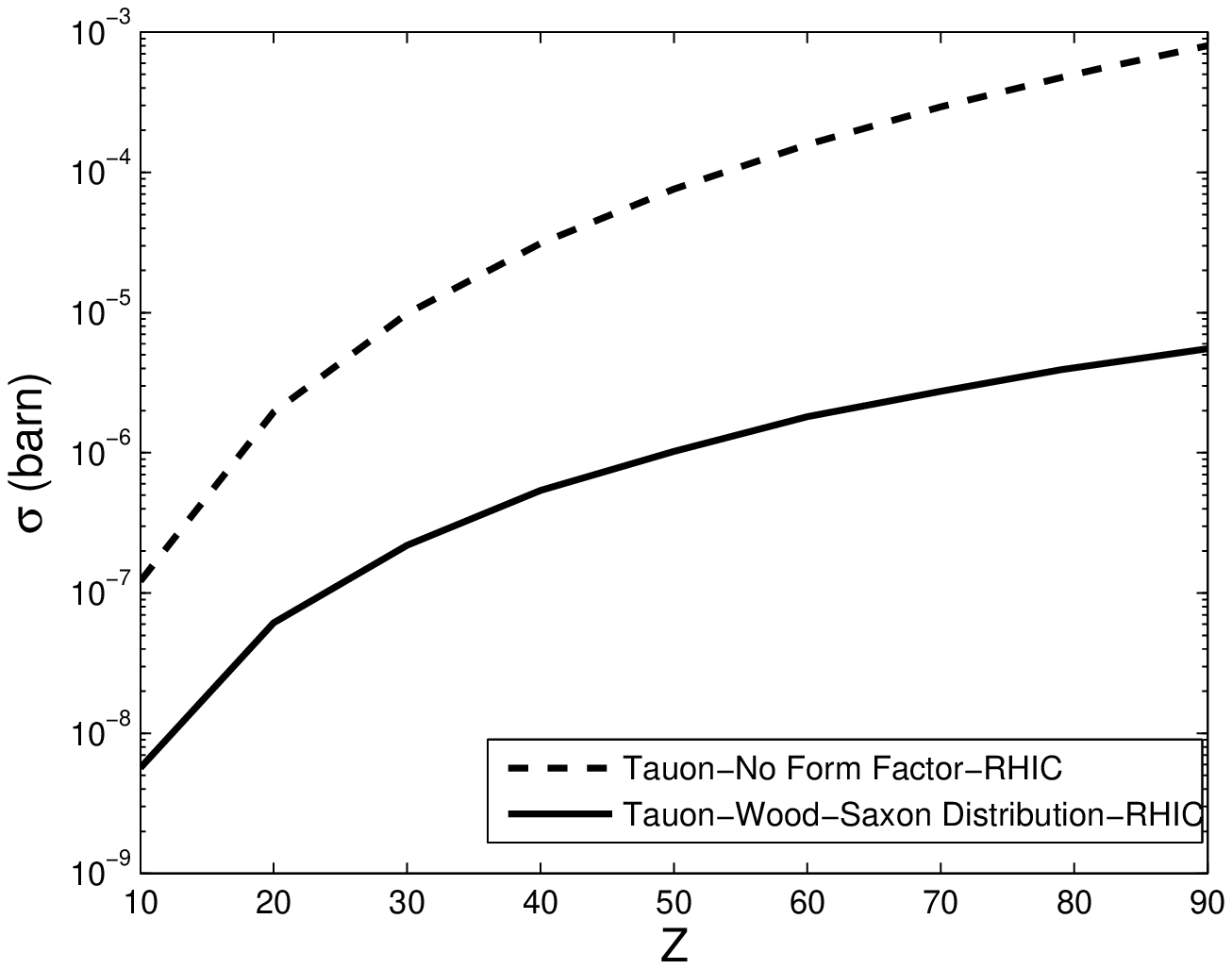}}               
  \subfloat[]{\includegraphics[width=5.5cm,height=4.5cm]{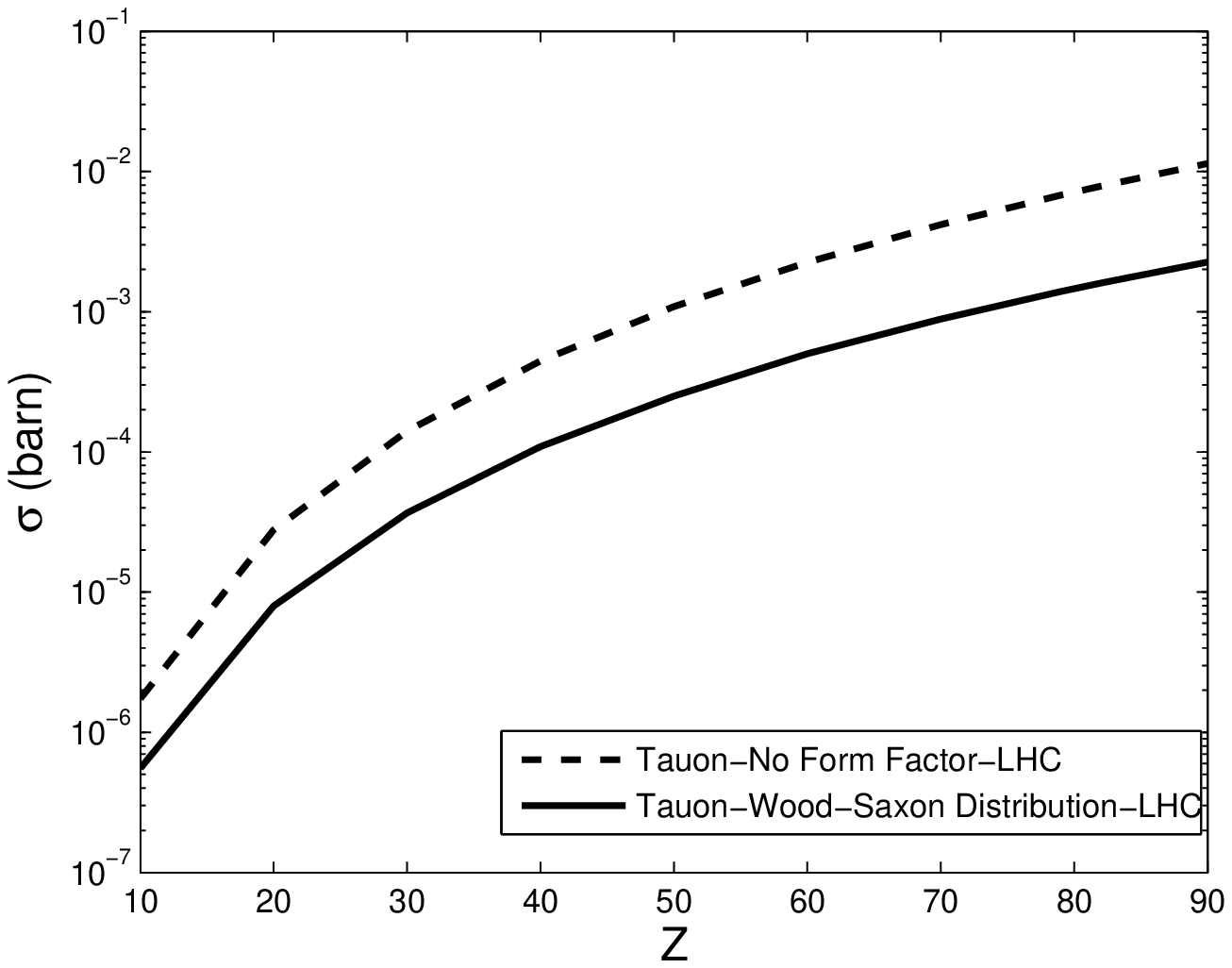}}
  \caption{Muon and tauon cross sections for RHIC and LHC energies as functions of the nuclear charge Z with and without form factors.}
\label{f26}
\end{figure}

Fig.~\ref{f26} shows the total cross section of heavy lepton pair production as a function of the charge of the colliding nuclei $Z$.
Obviously nuclear form factor reduces the total cross sections.
At RHIC energies, the form factor reduces the tauon pair production cross section about one or two order of magnitude, however
for the muon pair production the reduction is about two or three factors. On the other hand, at LHC energies, form factors reduce the muon pair production cross section about one or two factors, for the tauon production cross section this reduction is slightly less than one order of magnitude.

\begin{figure}
  \centering
	\subfloat[]{\includegraphics[width=5.5cm,height=4.5cm]{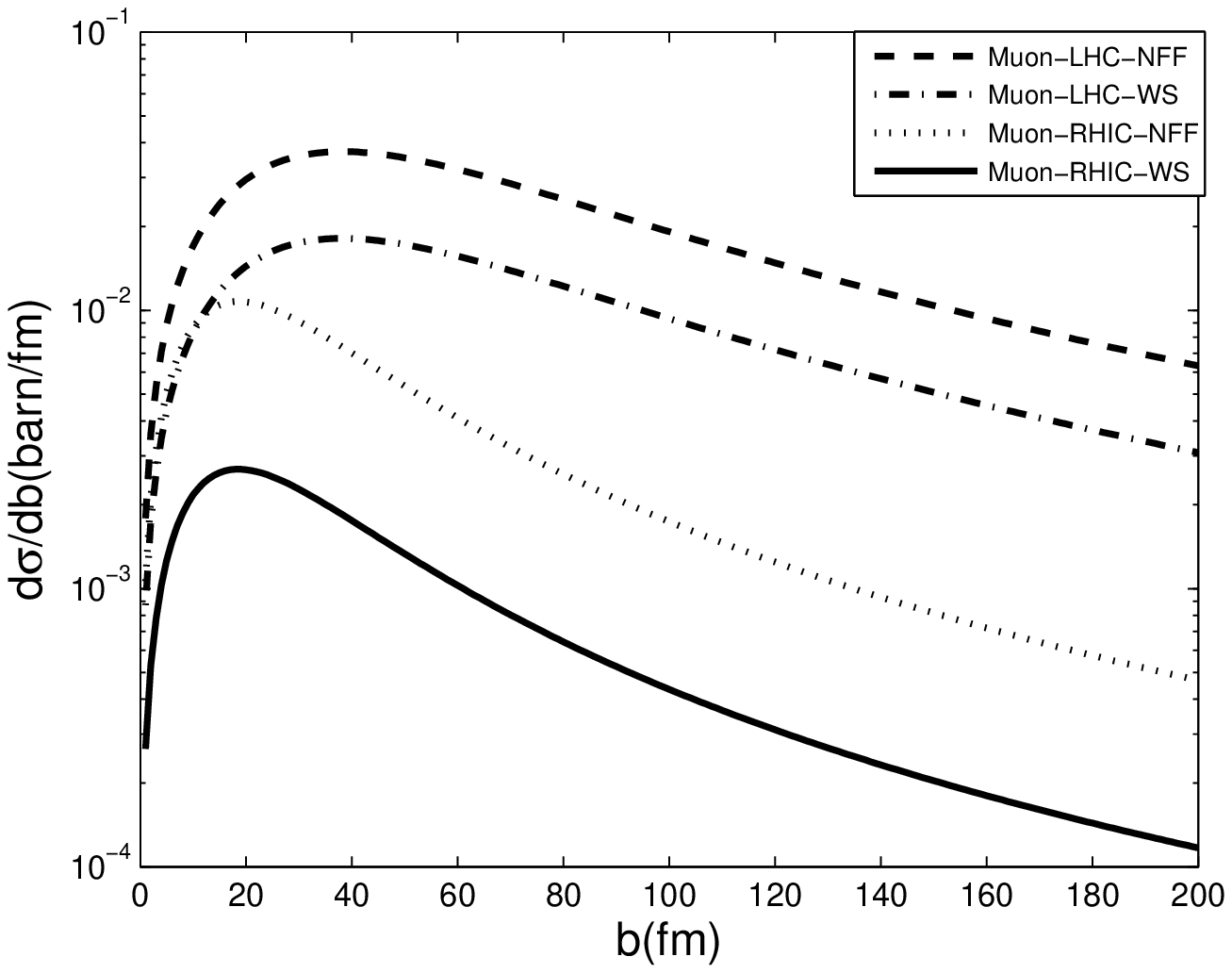}}
	\subfloat[]{\includegraphics[width=5.5cm,height=4.5cm]{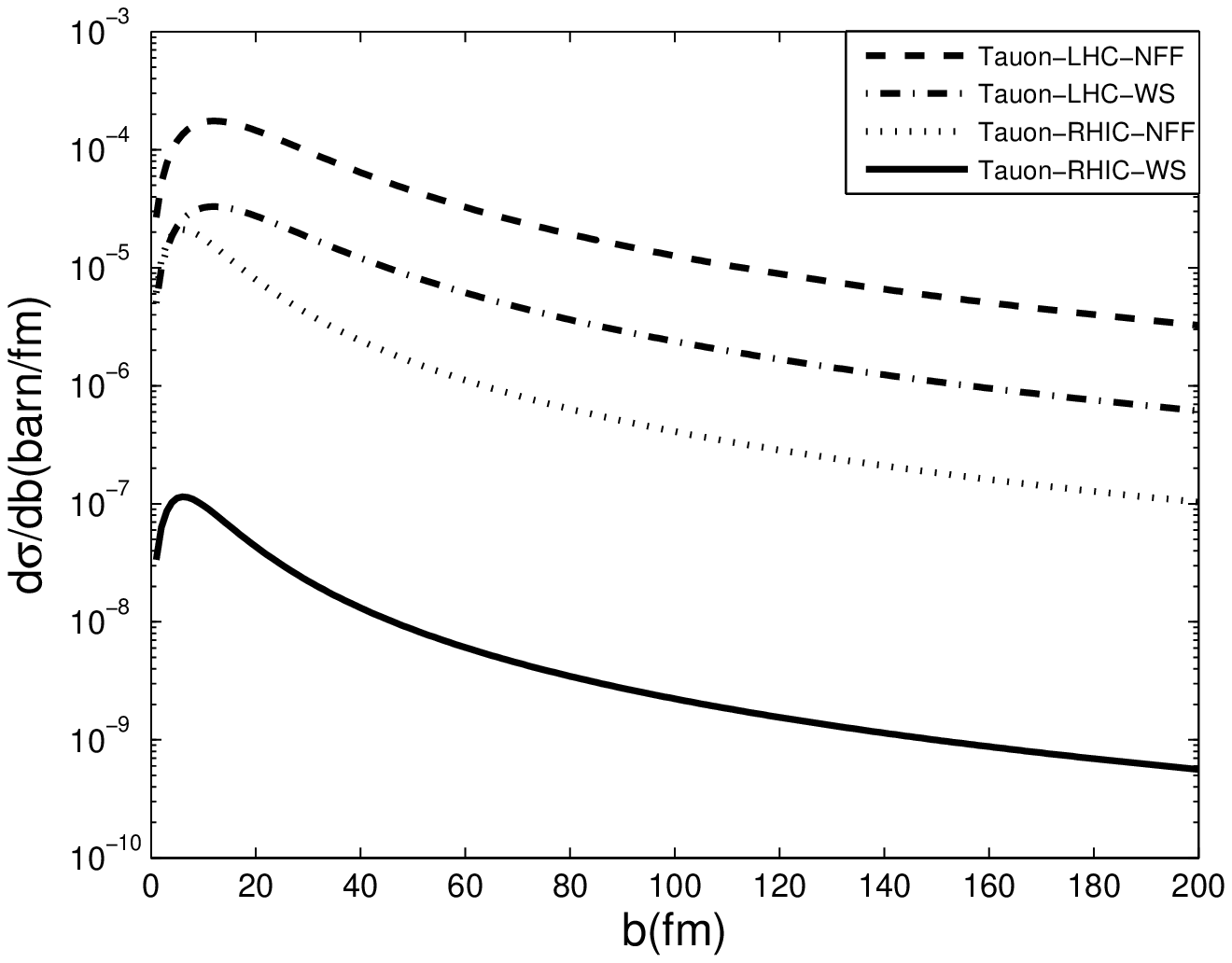}}\\
  \caption{Impact parameter dependence cross section for muon and tauon pairs at RHIC and LHC energies with Wood-Saxon distribution and no form factor.}
  \label{f27}
\end{figure}

Fig.~\ref{f27} displays the impact parameter dependence cross section for muon and tauon pairs production at RHIC and LHC energies with and without Wood-Saxon distribution.
In this calculations, the range of the impact parameter $b$ is between 0 and infinity. Since the Compton wavelengths of muon is $1.86 fm$ and tauon is $0.11 fm$ which are smaller than the radius of the nucleus, the heavy leptons are produced mainly near the radius of the heavy ions. Maximum production of muon pairs occurs at RHIC energies about impact parameters of 20 fm region, and at LHC energies in the impact parameter of 40 - 80 fm region. At about 200 fm region, the impact parameter dependence cross section decreases at about 10 factors at RHIC energies, and 2 factors at LHC energies.
On the other hand, maximum production of tauon pairs occurs about impact parameters of 7 fm region at RHIC energies, and about 10 fm region at LHC energies. At about 200 fm, the impact parameter dependence cross section decreases about 100 factors at RHIC energies and 50 factors at LHC energies.

Table~\ref{t2} represents the percentage of cross section contributions to the total cross section between the impact parameter regions $ b < R_{1}+R_{2} $ and $ b > R_{1}+R_{2} $ where $R_{1}+R_{2} \sim 15 fm$ for the heavy ions.  The cross section calculations are taken from the Table I with form factor. For the electron-positron pair production cross section, pairs are overwhelmingly produced for the region of $ b > R_{1}+R_{2} $. For the muon and tauon pair production, $ b < R_{1}+R_{2} $ region has important contribution to the total cross sections.
When the energy of the colliding ions increases, the contributions from the $ b < R_{1}+R_{2} $ region decreases. However, when the mass of the lepton increases, the contribution from the $ b < R_{1}+R_{2} $ region increases. We have also repeated the calculations for the impact parameters $b$ for the regions $20$, $25$ and $30$ fm.
This calculations clearly indicate that the details of the form factors, such as the distributions of the nucleons in the nucleus, are important for the calculations of the cross sections.

In all above calculations, we have also used proton form factor $G_{E}(q^{2})$. Our results show that the contribution of this proton form factor is negligible.
\begin{figure}[h]
  \centering
  \subfloat[]{\includegraphics[width=5.5cm,height=4.5cm]{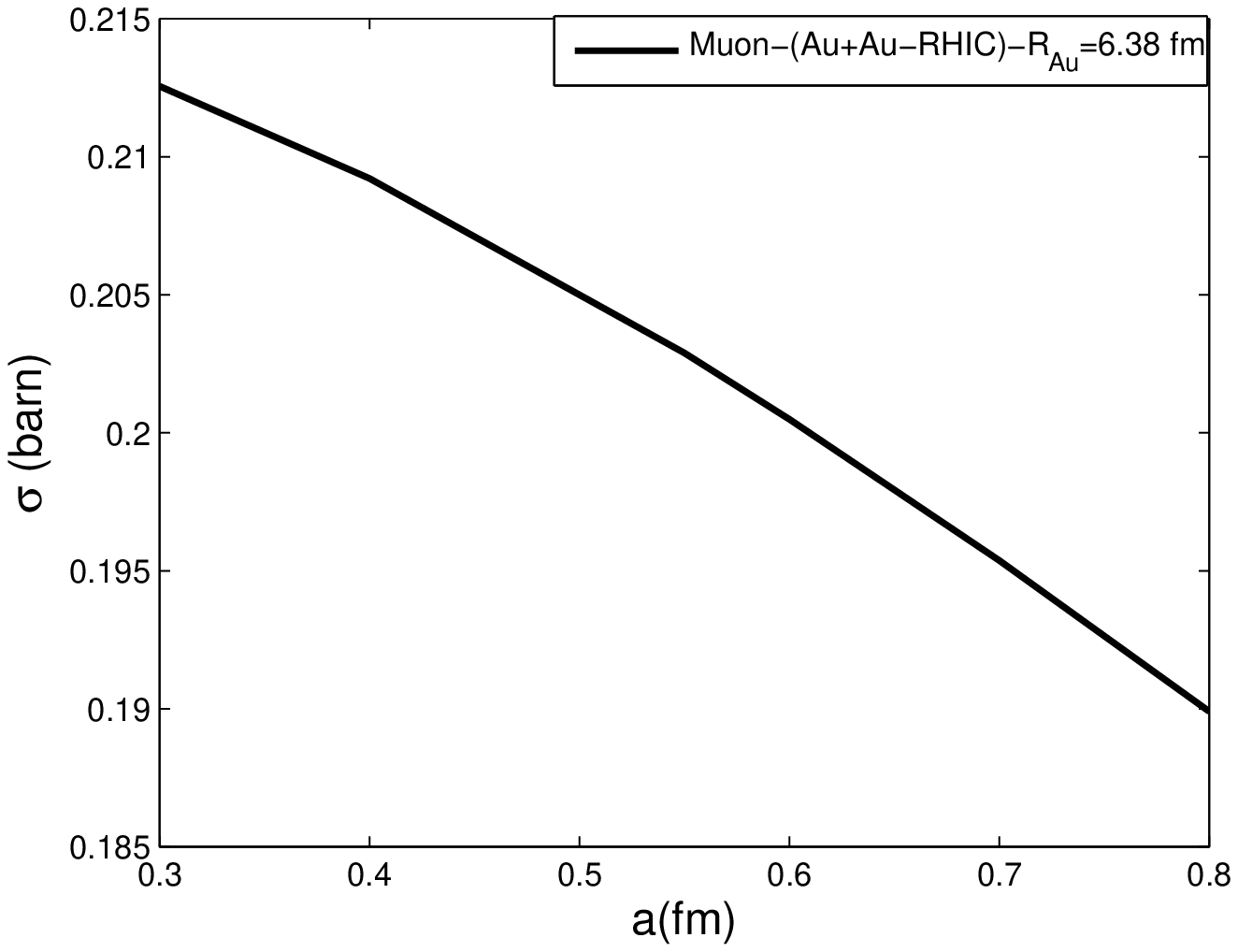}}                
  \subfloat[]{\includegraphics[width=5.5cm,height=4.5cm]{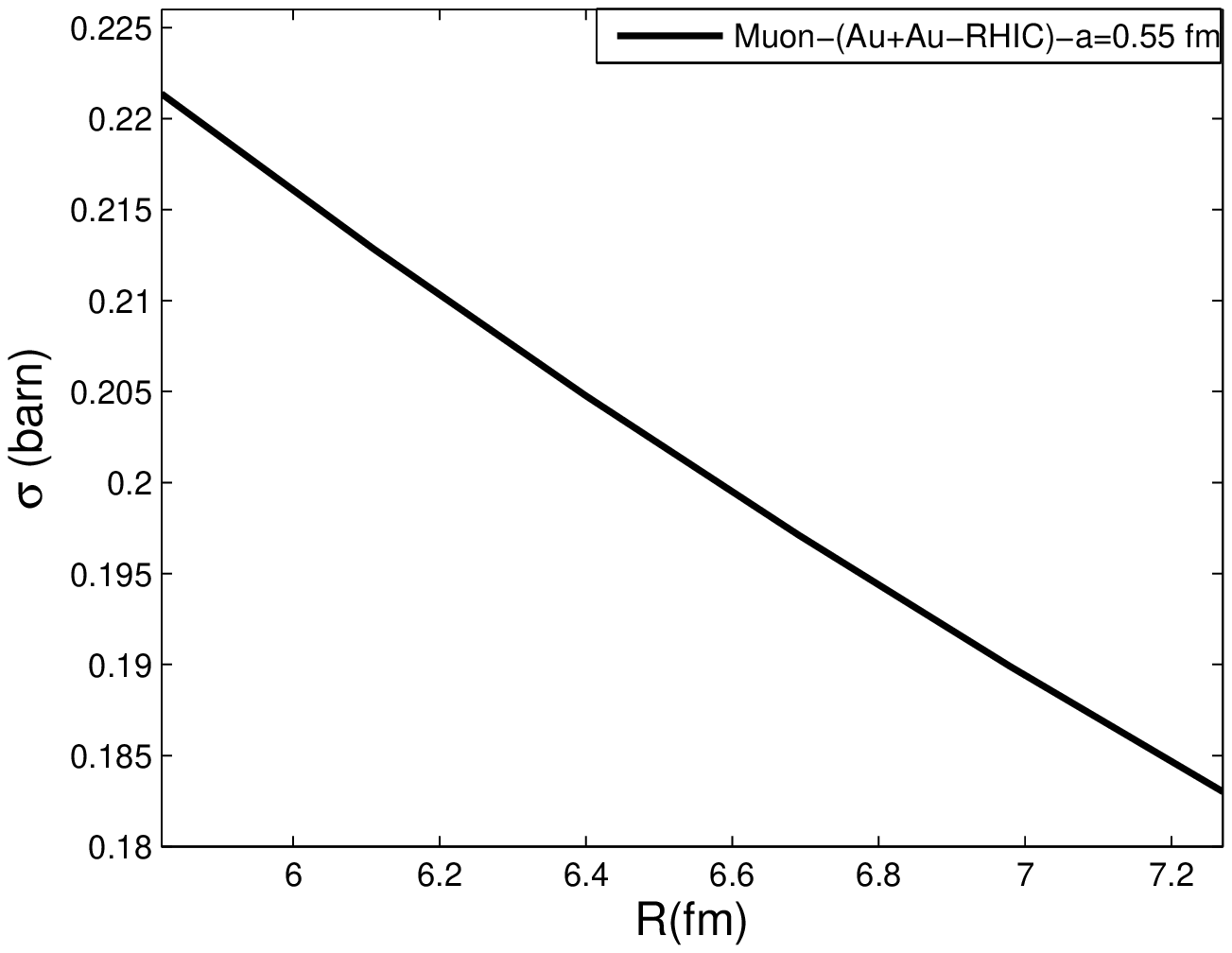}}\\
  \subfloat[]{\includegraphics[width=5.5cm,height=4.5cm]{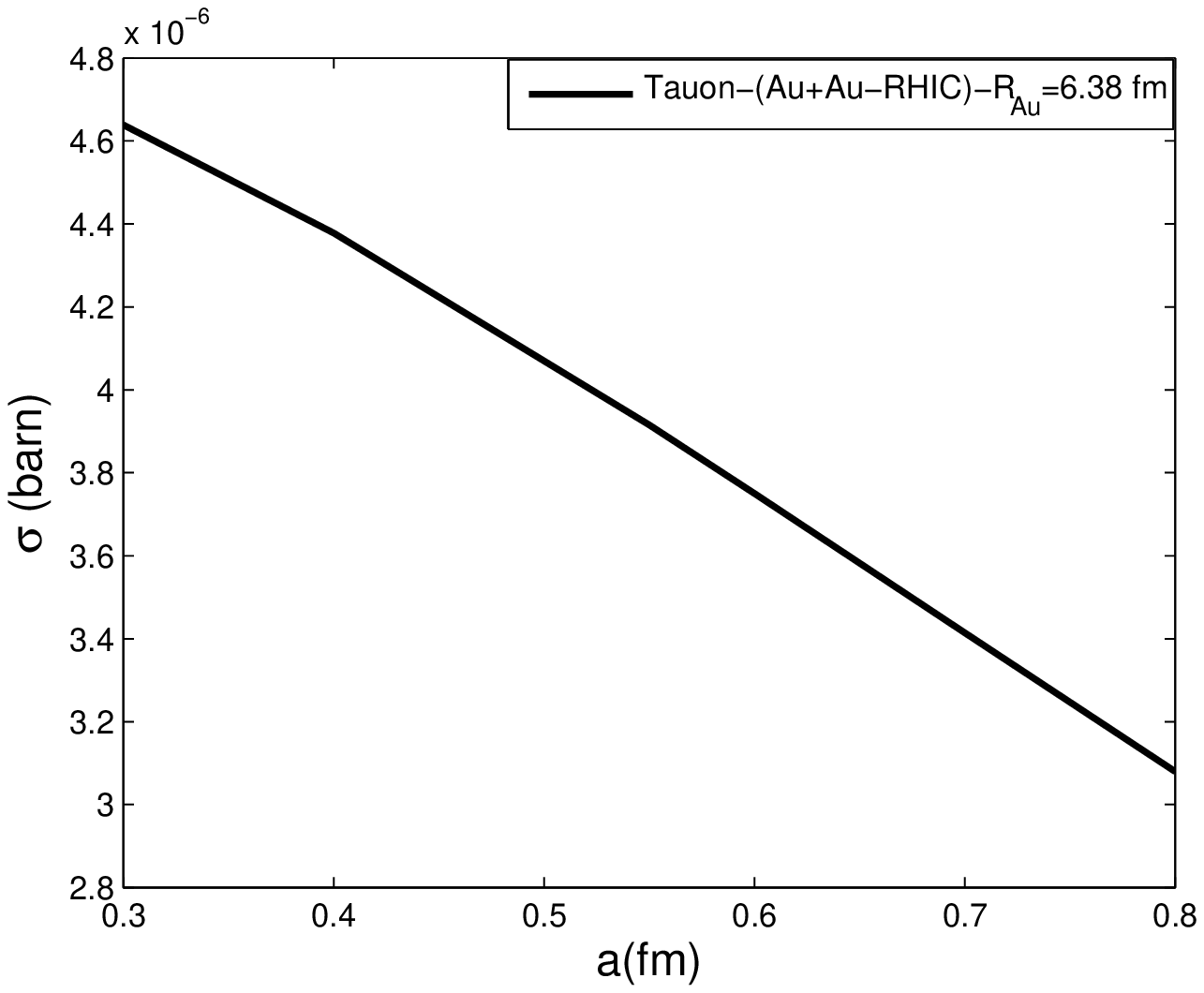}}               
  \subfloat[]{\includegraphics[width=5.5cm,height=4.5cm]{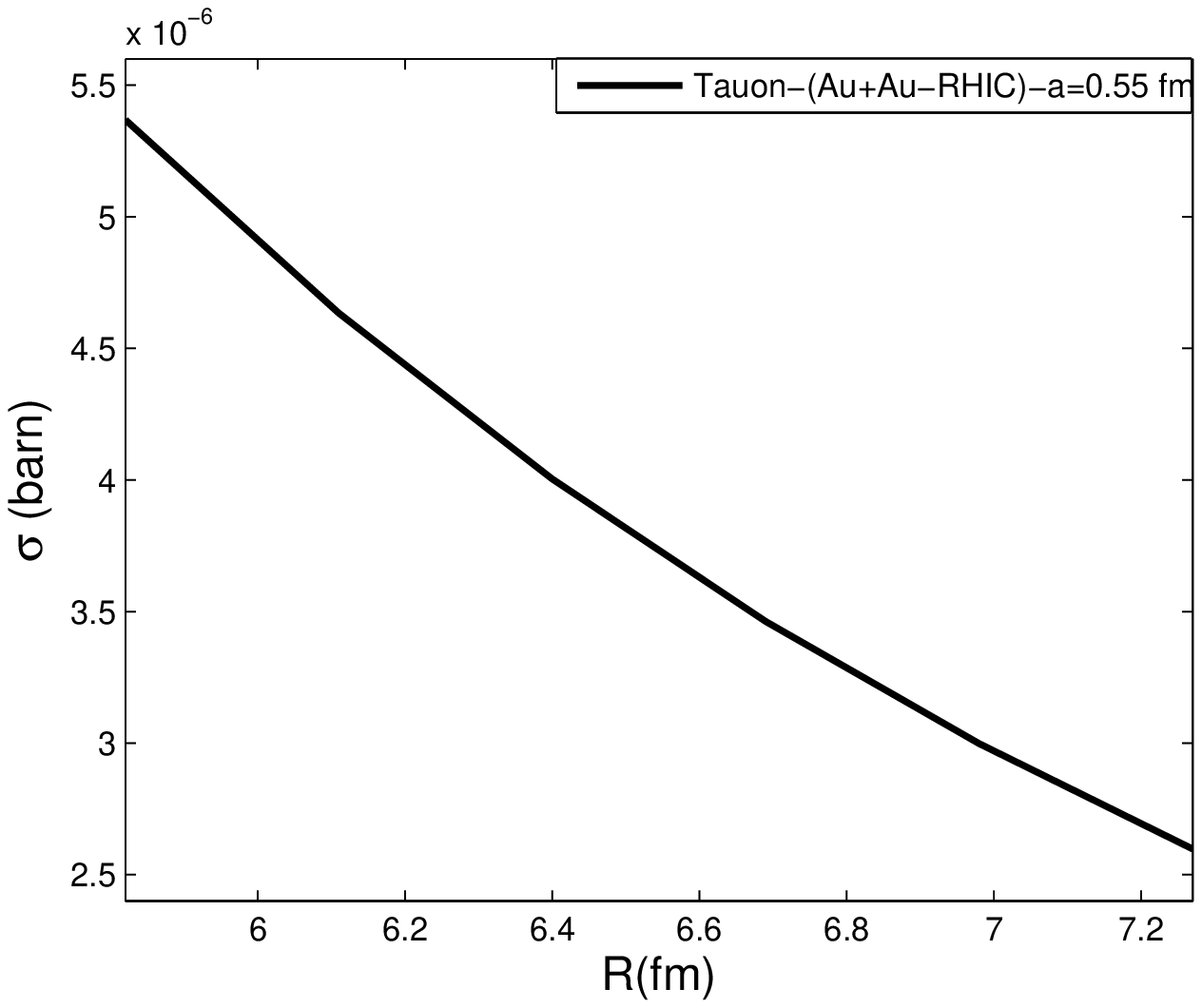}}
  \caption{ The cross section of produced muon a) as a function of skin thickness $a$ at RHIC energy when $R = 6.38 fm$ b) as a function of $R$ at RHIC energy when skin thickness $a=0.55 fm$. The cross section of produced tauon c) as a function of skin thickness $a$ at RHIC energy when $R = 6.38 fm$ d) as a function of $R$ at RHIC energy when skin thickness $a=0.55 fm$. }
\label{f28}
\end{figure}
\begin{figure}[h]
  \centering
  \subfloat[]{\includegraphics[width=5.5cm,height=4.5cm]{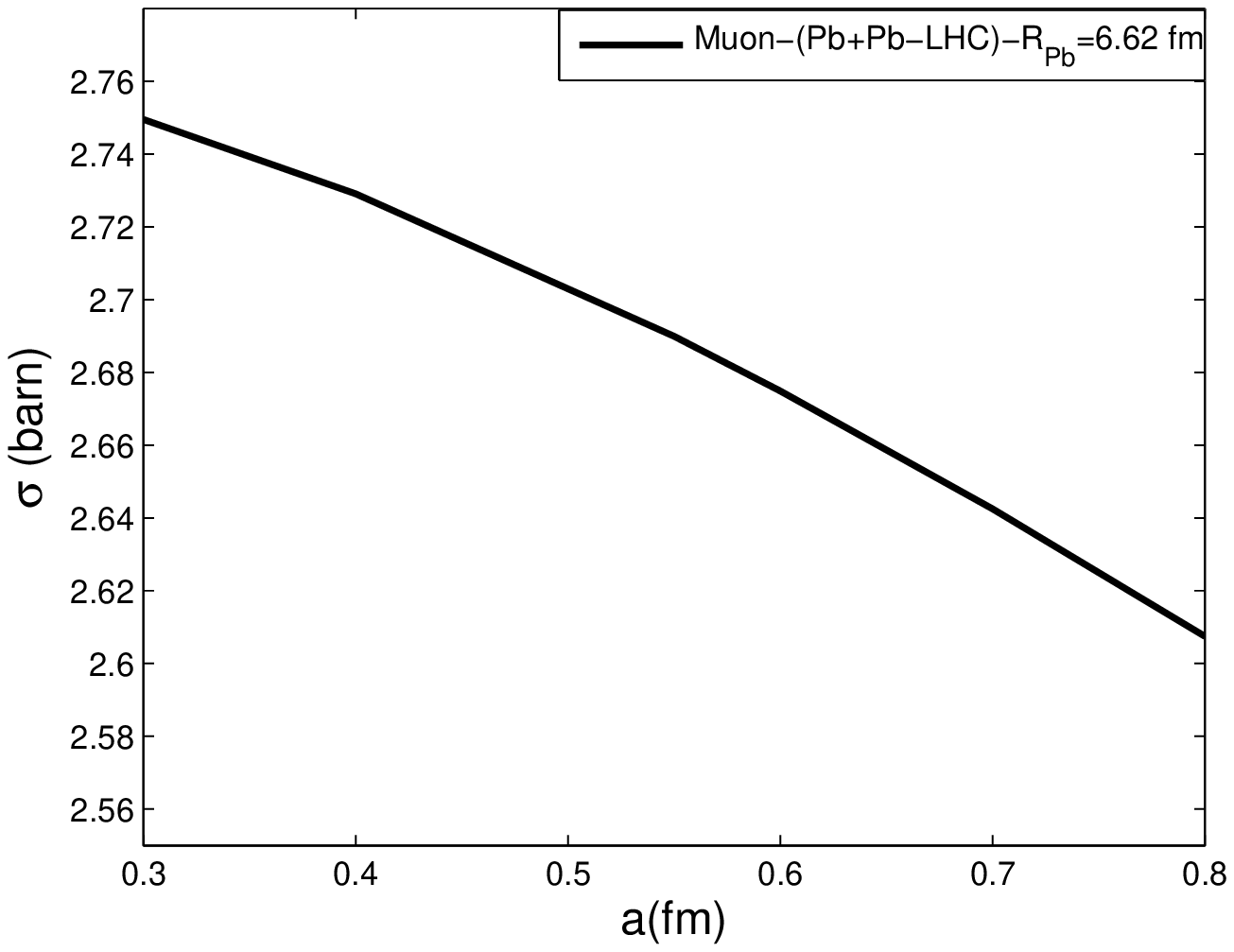}}                
  \subfloat[]{\includegraphics[width=5.5cm,height=4.5cm]{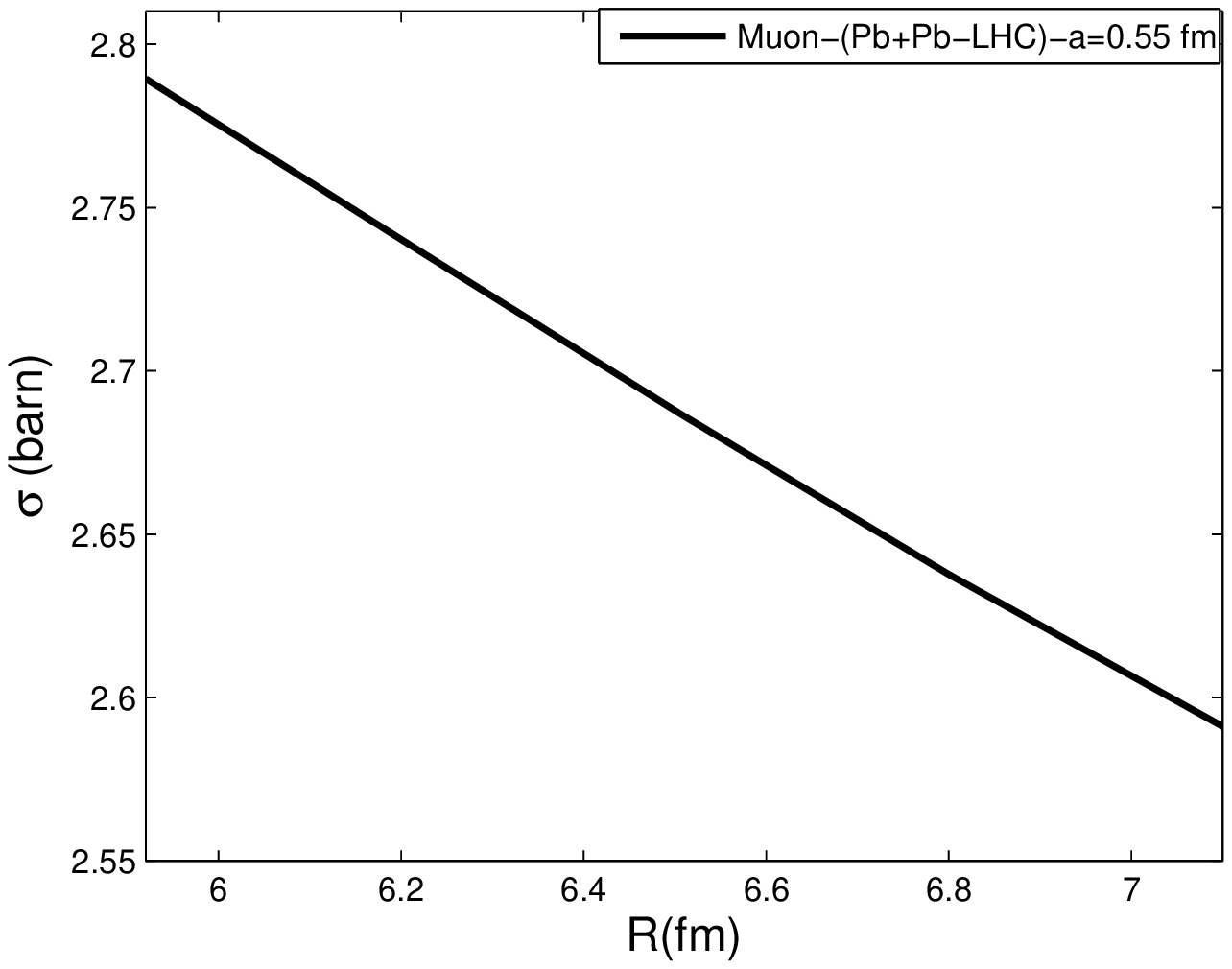}}\\
  \subfloat[]{\includegraphics[width=5.5cm,height=4.5cm]{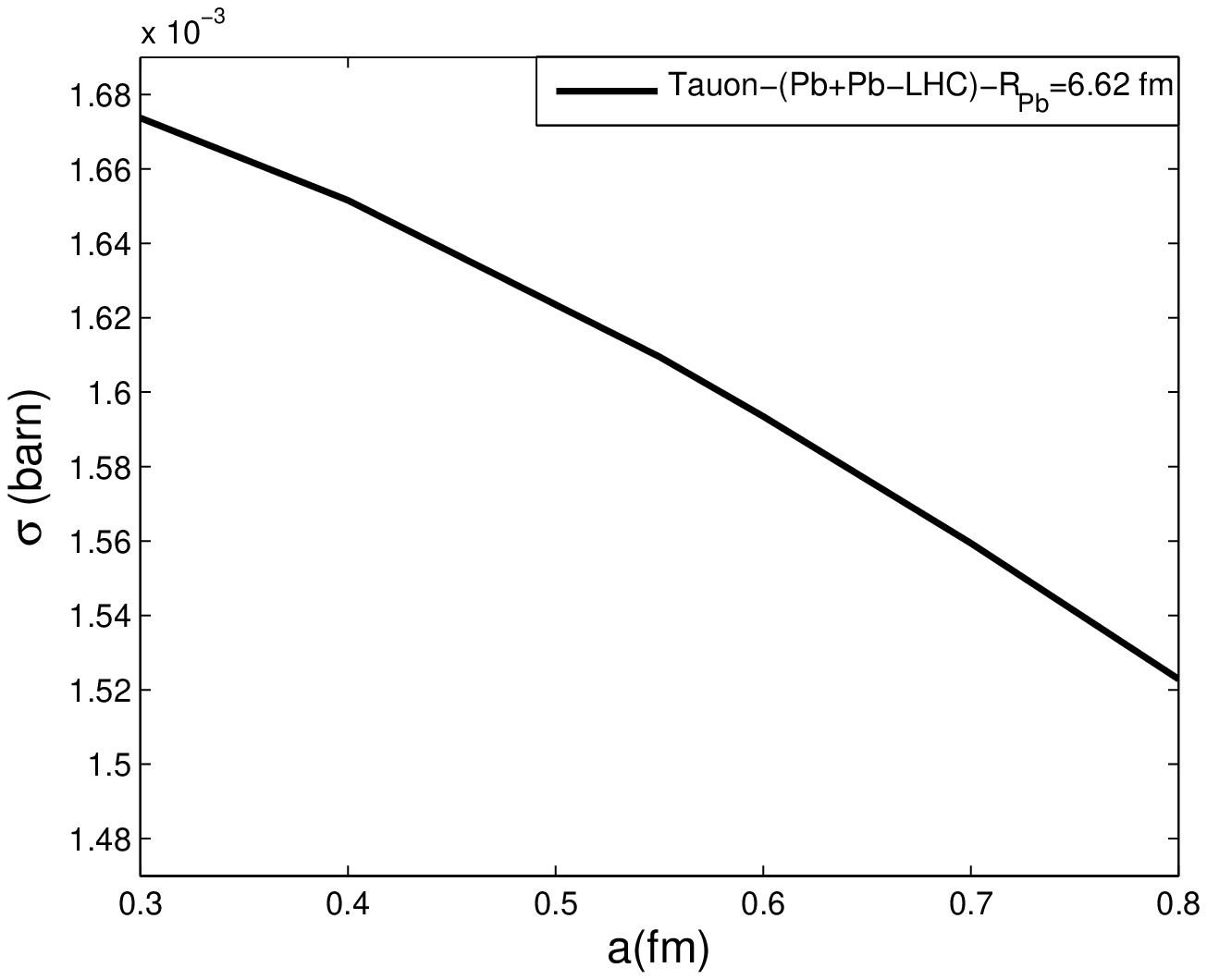}}               
  \subfloat[]{\includegraphics[width=5.5cm,height=4.5cm]{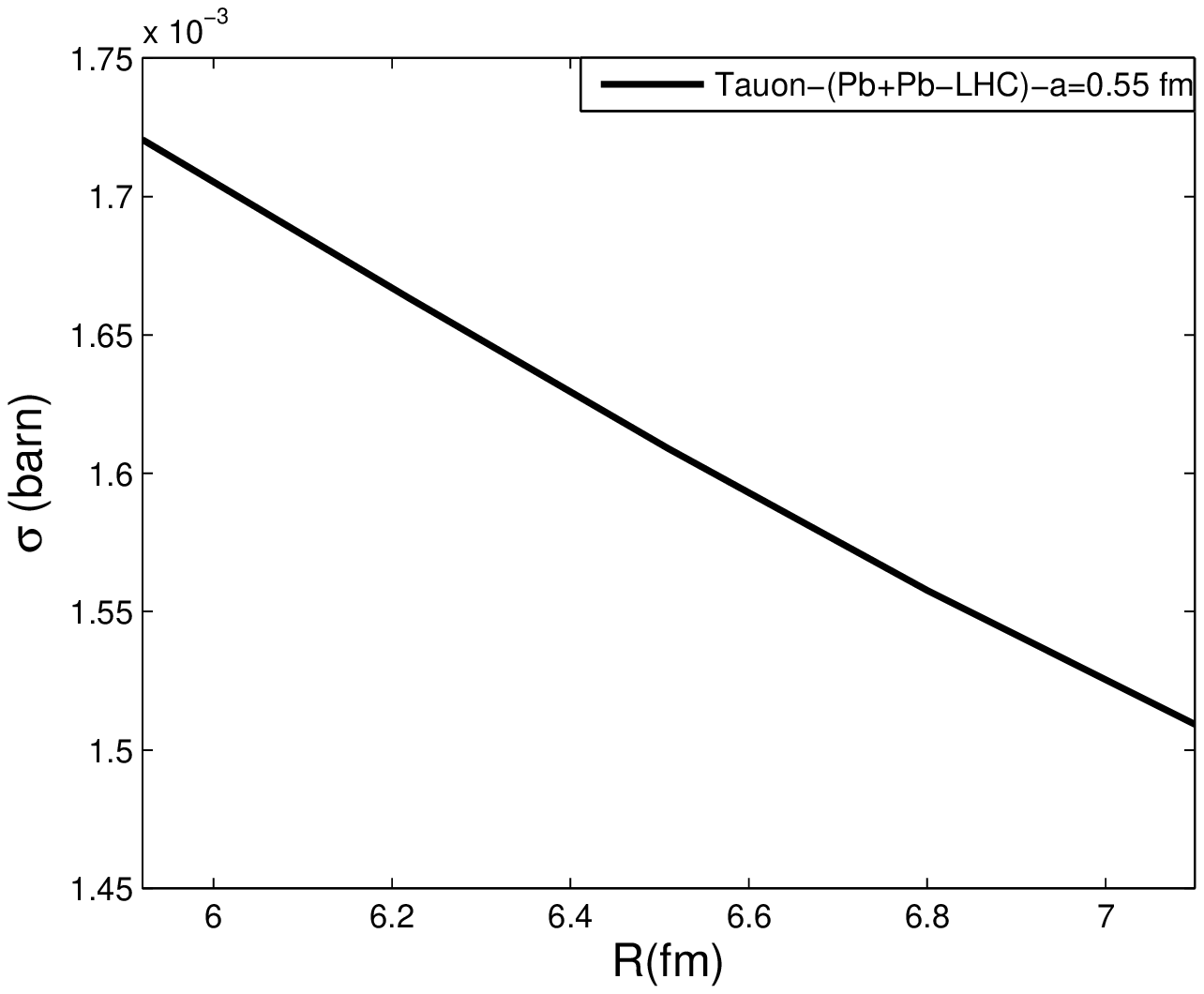}}
  \caption{ The cross section of produced muon a) as a function of skin thickness $a$ at LHC energy when $R = 6.62 fm$ b) as a function of $R$ at LHC energy when skin thickness $a=0.55 fm$. The cross section of produced tauon c) as a function of skin thickness $a$ at LHC energy when $R = 6.62 fm$ d) as a function of $R$ at LHC energy when skin thickness $a=0.55 fm$. }
\label{f29}
\end{figure}
The description of matter distribution in nuclei is still under investigation in nuclear physics and precise knowledge 
of this distribution can help us to understand the structure of the nucleus. Although the charge distribution has been measured with
high accuracy from the electron scattering data so that the charge radii are usually known with uncertainties
lower than 1$\%$ [1,2]. However, the neutron distribution in the nucleus is not accurately measured as protons.

We have also investigated the neutron distribution of the nucleus. Wood Saxon equation gives us the proton charge distribution and neutron density distribution of the nucleus. In Eq. 8, the parameter $ R $ is the radius of the nucleus and $a$ is the skin depth and these parameters are obtained from the electron scattering data \cite{barrett}. Recent experiments show that neutrons are distributed differently from the protons and this effects the cross sections of the heavy lepton pair productions. Neutrons are electrically neutral, so do not contribute to the electric form factor. Neutrons can contribute to the hadronic radius.
In order to see the effects of the neutron distribution, we have changed the parameters $R$ and $a$ in the Wood-Saxon form factor. First we keep the $R$ constant and change the parameter  $a$ between $ 0.3 fm\leq a \leq 0.8 fm$. Although the parameter $a$ has a value of about $0.55 fm$, by changing it for the extreme ranges, we can see how it effects the cross sections. 
Next we keep the parameter $a=0.55 fm$ constant and the change the radius $R$ between $5.8 fm \leq R \leq 7.2 fm $. In Figs.~\ref{f28} and~\ref{f29}, we have plotted these cross sections for muon and tauon productions at RHIC (gold nucleus) and at LHC (lead nucleus). 
When the parameter $a$ is changed from $0.4 fm$ to $0.6 fm$ value, the cross section ($\Delta \sigma /\sigma$) of the produced muon pairs decreased by about $ 4 \% $  at RHIC and $2 \%$ percent at LHC. Similarly, tauon production
decreased by about $4\%$ at LHC and $14\%$ at RHIC. On the other hand, when we change the radius $R$ from from $6.1 fm $ to $ 7.0 fm$, the cross sections of the muon pairs decreased by about $5\%$ at LHC and $11\%$ at RHIC. The similar calculation shows that tauon production decreased by about $10\%$ at LHC and $35\%$ at RHIC. These calculations are tabulated in Table III. From this calculations, we can conclude that Wood-Saxon distribution function is more sensitive to the parameter $R$ compare to the parameter $a$. It also shows that, cross sections are very sensitive to the parameter $R$ at RHIC energies compare to the LHC energies. The proton distribution in the nucleus is well known. 
If the neutron surface is shifted outwards the proton surface this means that the skin depths of neutrons and protons are 
nearly equal $a_{n} \sim a_{p}$ and it is called $bulk$ or $skin$ type of neutron skin. On the other hand, if the diffuseness 
of the neutron surface is greater than the proton then $R_{n} \sim R_{p}$ and it is called $surface$ or $halo$ type of the neutron skin \cite{centelles}.

We have compared our results with the works done in Ref.\cite{vidovic}. Both cross section results of the RHIC for $\mu^{-}\mu^{+}$ production  nearly agree each other, however $\tau^{-}\tau^{+}$ cross section in our work approximately is $3$ times larger than their results. Our LHC cross section results of $\mu^{-}\mu^{+}$ and $\tau^{-}\tau^{+}$ productions are slightly higher than their results. Authors in Ref.\cite{vidovic} they used
Gaussian distribution form factor and the equivalent photon approximation to calculate the heavy lepton pair production cross section. 

When we compare our cross section results with Ref.\cite{bottch1}, both calculations for $\mu^{-}\mu^{+}$ production at RHIC energies are almost similar with and without form factor 
At RHIC energies, $\tau^{-}\tau^{+}$ cross section with no form factor in our work approximately is $5$ times lower than in Ref.\cite{bottch1}, however the results with form factor is close to each other. 

Our cross section results of $\mu^{-}\mu^{+}$ rates at RHIC and LHC are about slightly lower than in Ref.\cite{henc1} that is used Born calculations at RHIC and LHC energies. 
Since the parameter $Z\alpha \approx 0.6$ is not small, Coulomb and unitarity corrections
\begin{eqnarray}
\sigma_{total} = \sigma_{two-photon} + \sigma_{Coulomb} + \sigma_{unitary}
\end{eqnarray}
should be included in the calculations to obtain exact cross sections. In Ref.\cite{henc1} authors show that for
electron positron pair production, the Coulomb corrections to the Born cross section are large, whereas the unitary corrections
are small. On the other hand, for muon pair production the Coulomb corrections to the Born cross section (corresponding to multiphoton
exchange of the produced muon pairs with the nuclei) are small, whereas the unitary corrections (corresponding to the
exchange of light-by-light blocks between nuclei) are large. Authors find that for $Pb + Pb$ collisions at LHC there is about $-50 \% $ unitarity corrections to the Born cross sections. 

\section{Results}\label{s3}
In this work, we investigate the contribution of the Wood-Saxon nuclear form factors on the heavy lepton pair production.
We study the pair productions for the RHIC and LHC energies for the Au and Pb nucleus. Even though electron 
pair production is not strongly effected by the form factors, the calculations 
show that nuclear form factors play important role for the muon and tauon pair production. At RHIC and LHC energies, muon 
pair production is reduced by about 3 and 2 factors by the Au and Pb form factors, respectively. On the other
hand, at RHIC and LHC energies, tauon pair production is reduced by about 100 and 5 factors by the Au and Pb form factors, 
respectively. These calculations show that tauon pair production is very sensitive to the form factors at the RHIC energies.
We can conclude that, muon pair production in peripheral heavy-ion collisions is coherent greater than the dimensions 
of the nucleus. However, although for the LHC energies the tauon pair production is coherent over distances of the
dimension of the nucleus, for the RHIC energies this coherence is break down. Since the mass of the tauon
is very large when compared to muon, incoherent pair production of tauon is dominant at the RHIC energies. When the energy is
increased, the coherence pair production is restored. We expect that this result is also valid for the other heavy particles.

We have also investigated the effect of the neutron distributions in the nucleus to the cross section of the heavy lepton pair production.
Neutron skin can be produced in two ways: First type is as a result of the shift of the neutron surface in relation to the proton one (the ”bulk” or "skin" type $a_{n} \sim a_{p}$), and the second type is due to the differences in the surface thickness of proton and neutron matter in a nucleus (the ”surface” or "halo" type $R_{n} \sim R_{p}$). In general, both types of mechanisms contribute to the creation of neutron skin.
We have changed the parameters $a$ and $R$ in the Wood-Saxon form factor, and our calculations show that  Wood-Saxon distribution function is more sensitive to the parameter $R$ compare to the parameter $a$.

\end{document}